\documentclass[prx,aps,nofootinbib,amssymb,twocolumn,superscriptaddress,10pt]{revtex4-2}

\usepackage[utf8]{inputenc}
\usepackage{graphicx,xcolor} 
\usepackage{amsmath}
\usepackage{braket}
\usepackage{float}
\usepackage{ulem}
\usepackage{longtable}
\usepackage{booktabs}
\definecolor{orange(ryb)}{HTML}{FFA500}
\definecolor{dodgerblue}{HTML}{1E90FF}
\definecolor{pinkerton}{HTML}{EC368D}
\definecolor{forest}{HTML}{6DD189}
\usepackage[colorlinks=true,citecolor=dodgerblue,linkcolor=dodgerblue,urlcolor=dodgerblue,pdftitle={TQCmoire}]{hyperref}

\begin{document}

\title{High-throughput discovery of moiré homobilayers guided by topology and energetics}

\author{Naoto Nakatsuji}
\affiliation{Department of Physics and Astronomy, Stony Brook University, Stony Brook, New York 11794, USA}
\author{Jennifer Cano}
\affiliation{Department of Physics and Astronomy, Stony Brook University, Stony Brook, New York 11794, USA}
\affiliation{Center for Computational Quantum Physics, Flatiron Institute, New York, New York 10010, USA}
\author{Valentin Cr\'epel}
\affiliation{Department of Physics, University of Toronto, 60 St. George Street, Toronto, ON, M5S 1A7 Canada}

\date{\today}

\begin{abstract}
Van der Waals heterostructures promise on-demand designer quantum phases through control of monolayer composition, stacking, twist angle, and external fields. 
Yet, experimental efforts have been narrowly focused, leaving much of this vast moir\'e landscape unexplored and potential promises unrealized. 
Here, we present a scalable workflow for high-throughput characterization of twisted homobilayers and apply it to $K$-valley semiconductors.
Combining small-scale density functional theory with perturbation theory, we efficiently extract moir\'e band gaps, valley Chern numbers, magic angles, and the threshold for lattice relaxation. 
Beyond this rapid high-throughput characterization, we
parameterize a continuum model for each material, which provides a starting point for more detailed study.
Our survey delivers an actionable map for systematic exploration of correlated and topological phases in moir\'e homobilayers, and identifies promising new platforms: chromium-based transition metal dichalcogenides for high-temperature quantum anomalous Hall effects, transition metal nitride halides for intertwined superconducting and moir\'e physics, and atomically thin III-V semiconductors for room-temperature-scale moir\'e effects.
\end{abstract}

\maketitle

\section{Introduction}

\begin{figure}[t]
\centering
\includegraphics[width=\columnwidth]{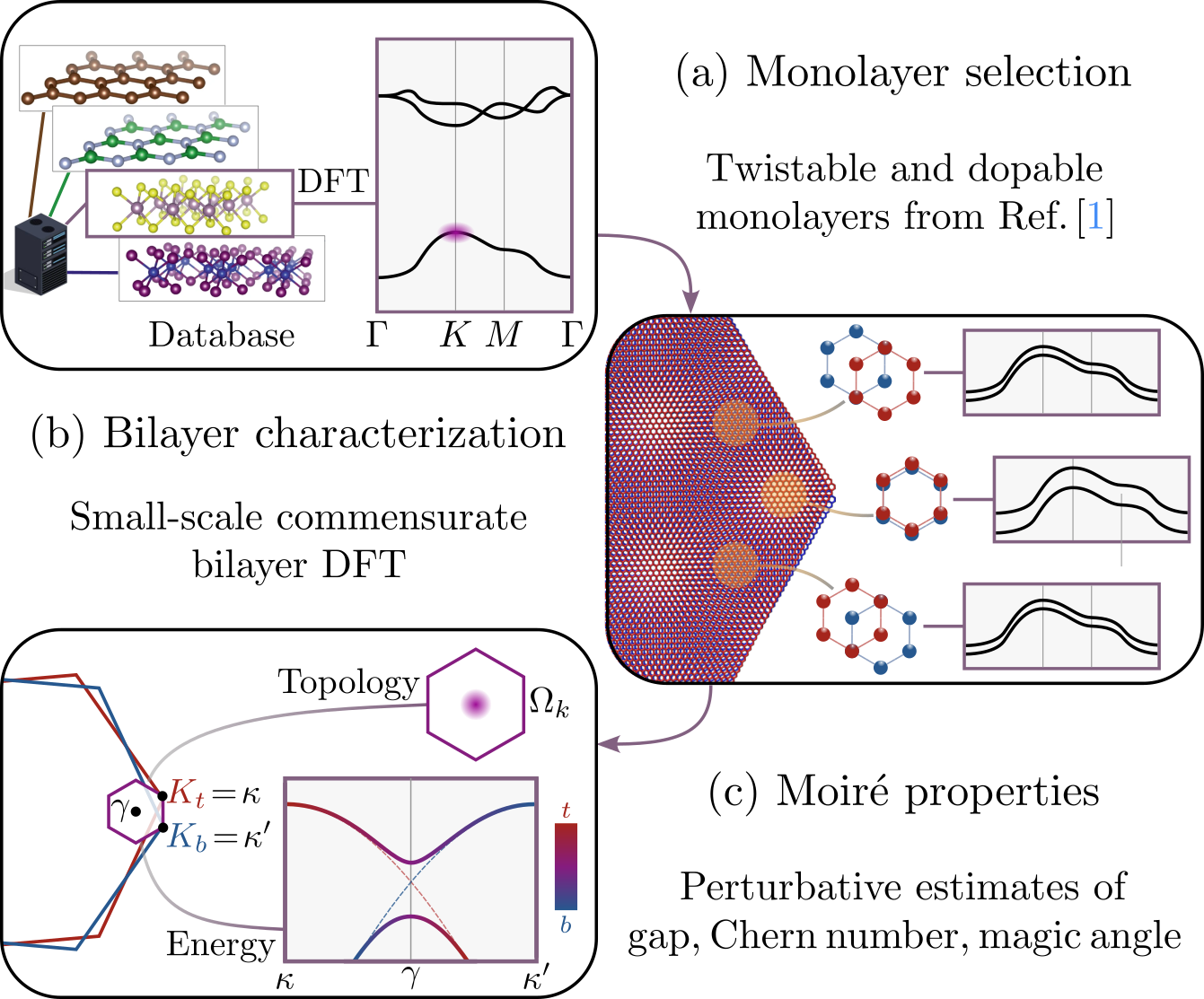}
\caption{\textit{
Schematic representation of our workflow for high-throughput characterization of twisted semiconducting homobilayers. (a) Starting from the list of twistable materials identified in Ref.~\cite{jiang20242d}, (b) small-scale commensurate DFT calculations efficiently characterize the moiré field variations across the long-wavelength moir\'e pattern, which (c) allows a rapid estimation of the valley-resolved Chern number, moiré band gap, and magic angle via perturbation theory~\cite{crepel2025efficient}. This perturbative limit is analytically justified when the moiré period is large compared to the monolayer lattice constant but small enough that the monolayer kinetic energy exceeds the moiré potentials and in-plane lattice relaxation remains weak, which typically holds for twist angles $\theta \gtrsim 3^\circ$.
}} 
\label{fig_workflow}
\end{figure} 

Van der Waals heterostructures -- assemblies of atomically thin two-dimensional materials -- have emerged as a cornerstone of quantum materials research~\cite{geim2013van,liu2016van,novoselov20162d,castellanos2022van}.
By stacking layers with controlled twist and alignment, these systems generate moiré superlattices that can host correlated and topological phases absent in their monolayer constituents~\cite{xiao2020moire,andrei2021marvels,kennes2021moire,mak2022semiconductor}. 
Landmark discoveries include superconductivity in twisted bilayer graphene~\cite{cao2018unconventional,lu2019superconductors,yankowitz2019tuning} and WSe$_2$~\cite{wang2020correlated,xia2025superconductivity,guo2025superconductivity,munoz2025twist,fischer2024theory}, as well as the realization of integer and fractional Chern insulators in twisted MoTe$_2$~\cite{cai2023signatures,xu2023observation,kang2024evidence,crepel2023anomalous,morales2023pressure,wang2024fractional} and BN-aligned rhombohedral graphene~\cite{lu2024fractional,xie2025tunable,waters2024interplay}. 
Despite these breakthroughs, the range of experimentally studied moiré systems remains surprisingly narrow, with most efforts concentrated on graphene and a limited subset of transition metal dichalcogenides (TMDs). 
This restricted focus stands in stark contrast to the expansive promise of moiré materials: the ability to design quantum phases on demand by tailoring layer composition, twist angle, and stacking geometry.

Realizing this vision requires a systematic understanding of how moiré design choices shape emergent electronic behaviors. 
This back-engineering has proved outstandingly challenging due to the limitations of conventional computational methods and the vast space of tunable parameters in these heterostructures~\cite{massatt2017electronic,carr2020electronic,chen2023moire,shaidu2025transferable}. 
Density functional theory (DFT), the standard approach for electronic structure, becomes prohibitively costly for moiré unit cells containing tens of thousands of atoms~\cite{zhang2022n} and struggles to capture their characteristic energy scales, which lie orders of magnitude below those for which DFT functionals have been optimized~\cite{verma2020status,hasan2025beyond}. 
Comprehensive DFT studies have therefore been limited to a few moir\'e materials~\cite{yananose2021chirality,lucignano2019crucial,jia2024moire,zhang2024polarization,zhu2025wavefunction}, often in response to experimental discoveries.
This reactive approach underscores the need for scalable and predictive frameworks capable of efficiently navigating the moiré parameter space and accelerating the search for new quantum materials.

Here, we introduce a high-throughput characterization of twisted $K$-valley homobilayers, which is both the largest class of twistable materials and the only one with experimentally demonstrated topological phases. 
Our workflow hinges on a rapid characterization of the moiré coupling geometry, enabling efficient extraction of the topological invariants, moiré band gaps, and magic angle for each bilayer using approximate methods whose regime of validity is assessed. 
Within this regime, the obtained parameters also define a continuum model for each bilayer, providing a starting point to more accurate simulations of the twisted heterostructures. 
Altogether, our results yield an actionable map of moiré materials to guide future experiments and more intensive calculations. 
In particular, our survey uncovers new platforms for exploring the interplay between topology and magnetism, with potential for robust anomalous Hall (QAH) effects; identifies candidates for room-temperature moiré physics; and reveals possibilities for moiré-induced high-$T_c$ superconductors.

\begin{figure*}
\centering
\includegraphics[width=\textwidth]{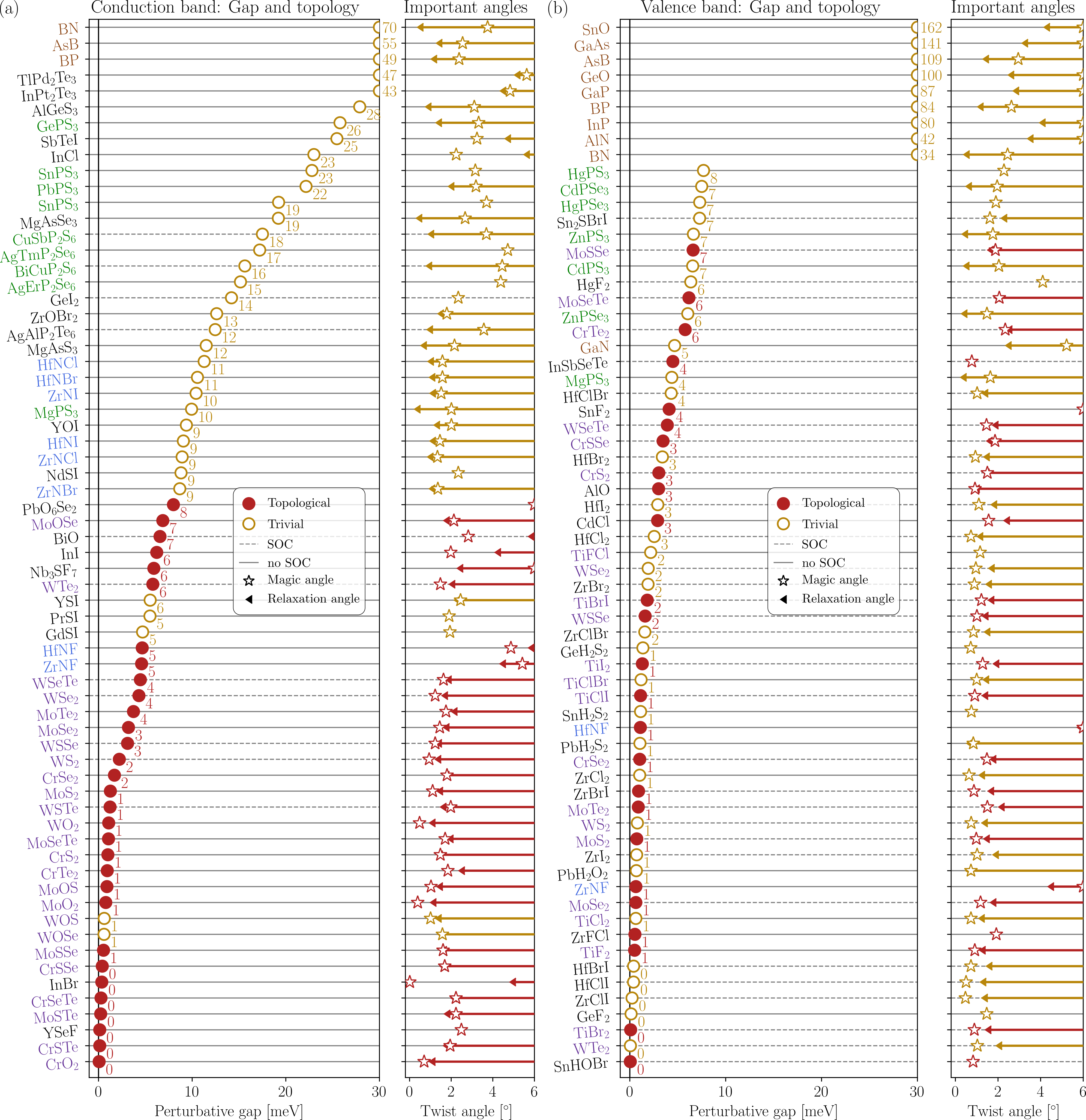}
\caption{\textit{
High-throughput characterization of moiré properties for all $K$-valley semiconducting moiré homobilayers. 
Each horizontal line corresponds to a theoretically stable twistable material with either (a) an electron-like or (b) a hole-like $K$-valley pocket~\cite{jiang20242d}; dashed (solid) gray lines indicates the presence (absence) of spin-orbit coupling. 
The left panels of (a) and (b) summarize the results of our perturbation theory~\cite{crepel2025efficient}: the position of the markers and the adjacent value gives the moiré gap amplitude (gaps $>30$meV are clipped on the edge) and their color encodes the topology of the bilayer (red for topological, gold for trivial).
The right panels show the estimated magic angles (stars) and twist angles $\theta_{\rm relax}$ where in-plane lattice relaxation becomes significant (triangles, none if elastic coefficients are not available). Our predictions are only formally justified above this relaxation threshold (thick colored line).
Names of certain material families are colored for reference in the text: transition metal dichalcogenides (purple), transition metal nitride halides (blue), transition metal phosphorus trichalcogenide (green), and a selection of one-atom thick componds (brown). 
The two lines labeled SnPS$_3$ represent different crystal structures with the same chemical composition; see full list of parameters in Methods.
}}
\label{fig_fullclassification}
\end{figure*}

\section{Workflow and method}

Our workflow for identifying promising moiré materials takes as input the list of twistable monolayers reported in Ref.~\cite{jiang20242d} (Fig.~\ref{fig_workflow}a), which performed a large-scale screening of entries in the Computational 2D Materials Database (C2DB) and Materials Cloud 2D Database (M2CD)~\cite{haastrup2018computational,mounet2018two,campi2023expansion,petralanda2024two} akin to other databases searches~\cite{ashton2017topology,zhou20192dmatpedia,crepel2023chiral,bao2024deep,kaplan2025}. 
To efficiently characterize the moiré coupling in the twisted homobilayers formed by these materials, we rely on two core principles. 
First, the local structure of the moiré coupling can be economically captured by a small set of commensurate DFT calculations~\cite{bistritzer2011moire,lei2025moire,jung2014ab,wu2019topological} (Fig.~\ref{fig_workflow}b), which bypasses ab-inito calculations for prohibitively large supercells. 
Second, several key physical properties -- including the moiré band gap, valley-resolved Chern numbers, and magic angle -- can be deduced from this coarse evaluation of the moiré fields~\cite{crepel2025efficient} (Fig.~\ref{fig_workflow}c), up to approximations justified in the regime where the moiré period is large compared to the monolayer lattice constant but small enough that in-plane lattice relaxation remains weak.
Although we illustrate the method for mirror-symmetric twisted homobilayers within a specific symmetry class, the workflow generalizes seamlessly to other lattice symmetries and moiré geometries, including heterobilayers and alternating trilayers. 
We now describe the three steps of our workflow (Fig.~\ref{fig_workflow}) in more details.

\paragraph{Symmetry-based monolayer selection:} 
The first stage of the workflow (Fig~\ref{fig_workflow}a) narrows the list of twistable materials by exploiting lattice symmetry, which is necessary for the derivation of topological symmetry indicators~\cite{bradlyn2017topological,po2017symmetry,cano2021band}. 
We start from the comprehensive set of 1647 twistable monolayers identified in Ref.~\cite{jiang20242d}, where ``twistable'' denotes materials that are theoretically stable (energy above the convex hull) and host Dirac cones or Fermi pockets that can be electrostatically doped. 
Among these, 1613 are semiconductors and only 34 are semimetals, making semiconductors the dominant class by almost two orders of magnitude, and the only one for which high-throughput screening is necessary.

We restrict the present study to twisted semiconductors formed from monolayers with hexagonal lattice symmetry and electrostatically dopable pockets at the $K/K'$ points. 
The choice of lattice symmetry selects the largest symmetry group among twistable semiconductors -- 55\% (887/1613) have three-fold symmetric lattices. 
The additional restriction to having $K/K'$-pockets (133/887) is mostly practical: to benchmark our predictions and topological classification against existing experimental data, we need to work in the material group containing twisted TMDs -- the only homobilayers where topological phases have been observed to date~\cite{cai2023signatures,xu2023observation,kang2024evidence}. 
The selected class of materials demonstrates our workflow's efficiency and scalability while providing direct validation against experiments. Future work will extend it to arbitrary lattices and types of Fermi pockets. 

\paragraph{Fast mapping of moir\'e parameters:} 
The second stage of the workflow economically extracts moiré couplings from small-scale commensurate DFT calculations (Fig.~\ref{fig_workflow}b). 
This strategy, pioneered early in the study of moiré materials~\cite{bistritzer2011moire,jung2014ab,lei2025moire}, exploits the fact that the long-wavelength moiré pattern locally resembles a commensurate bilayer with a relative shift between the layers. By computing the electronic structure of these shifted bilayers, we obtain local interlayer hybridization and intralayer potentials, which collectively determine the variation of these couplings across the moiré unit cell. 
Capturing the lowest harmonics of the moiré fields requires only three high-symmetry shifts. 
We further relate two of them by applying a mirror operation along the stacking axis ($M_z$) to one of the layers~\cite{wu2019topological}. 
This operation is trivial for monolayers that possess $M_z$-symmetry, but gives rise to two possible stackings for Janus materials; their difference is studied in Methods. 
Performing the remaining two commensurate DFT calculations using Quantum Espresso~\cite{giannozzi2009quantum}, we extract parameters encoding the lowest moiré harmonics, listed for all considered materials in Methods.
This dataset is valuable data in and of itself, as it parametrizes a leading-order continuum model for the twisted bilayers, a starting point for more detailed calculations including, for instance, many-body effects or lattice relaxation. 
Higher-order harmonics, which are relevant for the topology and energetics of higher moiré bands, can be obtained at the modest numerical overhead of more small-scale DFT runs.

Our atomic-scale commensurate bilayer calculations account for out-of-plane lattice relaxation (this effect may be enhanced at the moiré scale~\cite{wang2024fractional,jia2024moire}), but -- by construction -- cannot capture in-plane relaxations on the moiré scale. 
Our parameters therefore accurately describe the physics only when the latter remains negligible, an approximation that breaks down at very small twist angles~\cite{san2014electronic,nam2017lattice,carr2018relaxation}. 
To estimate the angle at which in-plane relaxation onsets, we extract the interlayer van der Waals energy scale $U_{\rm vdw}$ from step (b) and the Lamé elastic cofficient $\mu$ of the monolayer from C2DB~\cite{haastrup2018computational,campi2023expansion}. 
Together, these yield the relaxation angle $\theta_{\rm relax} \sim \sqrt{U_{\rm vdw} /\mu\eta }$~\cite{san2014electronic,nam2017lattice,carr2018relaxation,ezzi2024analytical,enaldiev2020stacking}, where $\eta=0.25$ is chosen so that atomic displacements at $\theta_{\rm relax}$ match those of $1.47^\circ$-twisted bilayer graphene~\cite{san2014electronic,nam2017lattice} or $1.85^\circ$-twisted WSe$_2$~\cite{carr2018relaxation,ezzi2024analytical}. 
Below $\theta_{\rm relax}$, renormalization of the moiré parameters renders our high-throughput predictions quantitatively inaccurate.
Nevertheless, these renormalizations can be incorporated into the continuum models of each bilayer using the local-stacking approximation~\cite{jung2014ab,koshino2020effective,nakatsuji2023multiscale,nakatsuji2025moir,zhang2025twist}, which requires the relaxation profile of the bilayer. While this lies beyond our high-throughput characterization, this profile can be obtained from the Lamé and $U_{\rm vdw}$ parameters~\cite{suzuura2002phonons,san2014electronic,nam2017lattice,ezzi2024analytical}, which we also provide in our parameter table in Methods.

\paragraph{High-throughput moiré characterization:} 
Finally, we use the DFT-extracted moiré harmonics to estimate several physical properties of the twisted bilayer, for which we characterize the regimes of validity (Fig.~\ref{fig_workflow}c).

Specifically, we compute the band gaps and topological character of the lowest energy moiré band with a symmetry-based framework~\cite{crepel2025efficient,lhachemi2025efficient,yang2024topological,liu2025symmetry}. 
To lowest order in the moiré field, the gaps that open at high-symmetry points of the moiré Brillouin zone (mBZ) are governed by how the leading moiré harmonics couple the monolayer wave functions~\cite{slater1952soluble}. 
The perturbative inclusion of these harmonics gives the gap and Bloch wave functions at the mBZ high-symmetry points, from which valley Chern numbers follow via symmetry indicators~\cite{fang2012bulk}. 
At twist angles where monolayer kinetic energy dominates over moiré coupling, this perturbative method is analytically justified. Fig.~\ref{fig_workflow}c summarizes this process: moiré couplings hybridize the bands from each layer and open a gap at the mBZ center ($\gamma$), which -- depending on the moiré geometry -- can generate Berry curvature and a nonzero Chern number.

The DFT-extracted moiré parameters also allow us to identify possible magic angles, where the bandwidth of the first moiré band reaches a local minimum as a function of twist angle, creating a fertile environment for correlated physics. 
To estimate these angles in a high-throughput manner, we extend the perturbative method to the mBZ corners ($\kappa/\kappa'$) and define the magic angle as the point where their energies equal that at $\gamma$. At this point, the lowest moiré band is constrained to interpolate between identical values, minimizing its bandwidth. Equivalently, this corresponds to the switch of band extrema from $\gamma$ to $\kappa/\kappa'$.

\section{Results and discussion}

\begin{figure*}
\centering
\includegraphics[width=\textwidth]{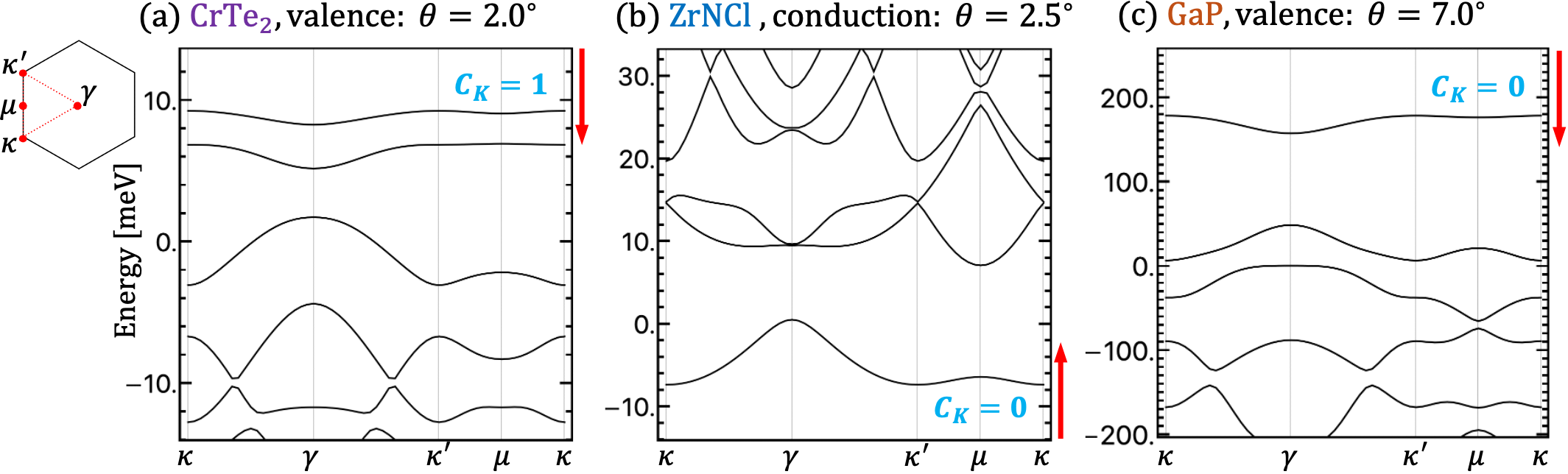}
\caption{\textit{
Moiré band structures for the continuum model of (a) hole-doped CrTe$_2$, a magnetic topological insulator; (b) eletron-doped ZrNCl, a 2d superconductor; and (c) hole-doped GaP, a large-angle moir\'e material. Their twist angles are $\theta = 2.0^\circ$, $2.5^\circ$, and $7.0^\circ$, respectively. The blue integers indicate the valley Chern numbers, the red arrow shows the doping direction. The inset in the upper-left shows the moiré Brillouin zone.
}}
\label{fig_bands}
\end{figure*}

Fig.~\ref{fig_fullclassification} summarizes our results, reporting the moiré gap amplitude and topological character of the lowest moiré band for all $K$-valley semiconducting homobilayers, along with their estimated magic angles and relaxation thresholds $\theta_{\rm relax}$. 
As an initial validation, our method predicts non-trivial topology and a gap of 1.1meV ($\simeq13$K) for twisted MoTe$_2$ in the valence band, consistent with experimentally measured Chern numbers~\cite{cai2023signatures} and the estimated activation gaps of $\sim 20–30$K at $3.9^\circ$~\cite{redekop2024direct,park2025observation}. Minor quantitative discrepancies are expected due to stronger out-of-plane lattice relaxation at the moiré scale~\cite{wang2024fractional,jia2024moire}, and the approximate methods employed in our high-throughput workflow (noting however that $3.9^\circ$ lies safely above $\theta_{\rm relax}$). 
This experimetal validation asserts the reliability and predictive power of our approach, which we now leverage to identify promising candidate moiré materials.

\paragraph*{Wealth of topological bilayers --- } 

A striking outcome of our screening is that nearly half (42\%, red symbols in Fig.~\ref{fig_fullclassification}) of all $K$-valley homobilayers exhibit topological moiré bands at angles. This shows that topological moiré physics is far more widespread than previously recognized and uncovers new candidate materials. 
For instance, while TMDs (MXY with ${\rm M} \in \{ {\rm Ti}, {\rm Cr}, {\rm Mo}, {\rm W} \}$ and ${\rm X}, {\rm Y} \in \{ {\rm O}, {\rm S}, {\rm Se}, {\rm Te} \}$, purple labels in Fig.~\ref{fig_fullclassification}) dominate experimental efforts~\cite{mak2022semiconductor,nuckolls2024microscopic}, they generally yield relatively small moiré gaps in the conduction band. 
By contrast, our study identifies material families with topological gaps many times larger, paving the way for robust topological moiré phases. In particular, HfNF and ZrNF, fluorine-based transition metal nitride halides (TMNHs, with ${\rm M} \in \{ {\rm Zr}, {\rm Hf} \}$ and ${\rm X} \in \{ {\rm F}, {\rm Cl} , {\rm Br} , {\rm I} \}$, blue labels in Fig.~\ref{fig_fullclassification}) exhibit nontrivial gaps greater than most TMDs. 
Among materials without spin–orbit coupling, these compounds offer some of the largest topological gaps, realizing robust $\mathbb{Z}_2$ topological insulators at full filling.

As previously discussed, the results in Fig.~\ref{fig_fullclassification} accurately capture bilayer physics for twist angles above $\theta_{\rm relax}$, where in-plane lattice relaxation remains weak. 
This limitation does not imply that relaxation effects necessarily suppress topological phases. 
On the contrary, when incorporated into the continuum model, these effects can drive bilayers from a trivial phase at large twist angles into topological phases at smaller angles, through renormalization of the leading moiré harmonics. 
While such regimes lie beyond the scope of our high-throughput approach, they are accessible through more detailed modeling. 
We provide instructions for performing such calculations in Methods, where we showcase ZrBr$_2$,a compound where lattice relaxation drives a transition to a topological phase below $\sim 1.5^\circ$.

\paragraph*{Moiré and magnetism: high-temperature quantum anomalous Hall --- } 

Within the TMD family, the Mo-based Janus compounds and chromium-based materials stand out for their large topological gap. 
Beyond a topological band gap, the QAH effect requires spontaneous breaking of time-reversal symmetry by interactions, which can happen on two distinct scales: moiré or atomic. 
In twisted MoTe$_2$, ferromagnetism emerges from interactions between doped charges at the moiré scale~\cite{crepel2023anomalous,chen2025quantum,anderson2023programming}, limiting the temperature range to the energy of spin-flip moiré magnons ($\lesssim 1$meV)~\cite{gonccalves2025spinless}. 
By contrast, electrons in intrinsically magnetic materials polarize through couplings to atomic spins, achieving much higher ordering temperature. 
This strategy of combining topology with intrinsic magnetism, which proved successful in magnetic topological insulators~\cite{chang2013experimental}, makes chromium TMDs particularly compelling for high-temperature QAH. 
Indeed, they exhibit intrinsic ferromagnetism that persists down to the few-layer limit and survives near room temperature~\cite{van1980crse2,li2021van,zhang2021room,xian2022spin,wu2022plane,xiao2022novel,fan2024progress}. 
This can push the magnon energy far above moiré energy scales, enabling QAH behaviors limited by the moiré band gap, approximately $6$meV $\simeq 70$K for CrTe$_2$ (see Fig.~\ref{fig_fullclassification}), approaching liquid-nitrogen temperatures.

\paragraph*{Platforms for intertwined moiré and superconducting physics --- } 

The TMNH family, previously noted above for the large topological gap of its fluorine-based members, also includes the non-topological chlorine-based compounds ZrNCl and HfNCl, which we propose as unique platforms to explore the competition and interplay between moiré physics and superconductivity. 
Indeed, these monolayers exhibit superconductivity with two particularly desirable characteristics: ($i$) relatively high critical temperatures ($\sim 5-20{\rm K}$)~\cite{hase1999electronic,felser1999electronic,kasahara2015unconventional,nakagawa2021gate,crepel2022spin} comparable to the moiré coupling in the conduction band of the corresponding bilayers (see Fig.~\ref{fig_fullclassification}), and ($ii$) the emergence of superconductivity at extremely low doping concentration -- down to a carrier density of $0.0076$ charge per unit cell~\cite{nakagawa2021gate}, equivalent to a 2d density of $\sim 7 \cdot 10^{12} {\rm cm}^{-2}$, achievable by electrostatic gating.


Moiré couplings in the bilayer redistributes the density of states (DOS). 
By tuning twist angle and carrier density, the Fermi level DOS can be substantially enhanced, which in turn influences superconductivity.
For example, the strong gap opening near the $\mu$ point of the mBZ in $2.5^\circ$-twisted ZrNCl (Fig.~\ref{fig_bands}b) could increase the DOS twofold, potentially boosting the monolayer critical temperature to $\sim 40$K. 
Additional control via displacement fields could create regimes where only one layer is doped, subject to both intrinsic superconductivity and an imposed moiré potential of comparable strength -- a scenario reminiscent of rhombohedral graphene multilayers, where superconductivity is supressed under a weak superlattice induced by an aligned hBN substrate~\cite{lu2024fractional,xie2025tunable,waters2024interplay,zhou2021superconductivity,han2025signatures,choi2025superconductivity}. 
TMNH systems thus provide a promising platform to address fundamental questions about the competition and coexistence of moiré physics and superconductivity, as well as the possibility of twisted nodal superconductors~\cite{pixley2025twisted,zhou2023non,can2021high,volkov2023current,lee2021twisted,zhao2023time} or twist-angle tunable Josephson junction~\cite{yabuki2016supercurrent,farrar2021superconducting,jian2022superconducting,tani2025twist}.

\paragraph*{Moiré physics at room temperature --- } Finally, we highlight the materials with the strongest moiré couplings identified in our search. 
The first major class comprises atomically thin binary hexagonal lattices, including III–V semiconductors (BN, BP, GaP, GaAs, \textit{etc}) and metal oxides (SnO, GeO), shown in brown in Fig.~\ref{fig_fullclassification}. 
Their perfectly two-dimensional structure and small interlayer spacing of about $3\text{\AA}$ enable strong tunneling -- up to $\sim 100$meV -- producing extremely narrow bands even at large twist angles, as illustrated in Fig.~\ref{fig_bands}c for $7^\circ$-twisted GaP. 
Achieving flat bands at such angles reduces sensitivity to disorder, such as twist-angle variations or strain, due to the smaller size of the moiré unit cell~\cite{carr2018relaxation,mcgilly2020visualization,benschop2021measuring,PhysRevB.105.245408,naik2022twister,tilak2022moire,cazeaux2023relaxation,kapfer2023programming,nakatsuji2023multiscale,redekop2024direct,hoke2024imaging,lai2025moire,crepel2025topologically,de2025theory}. 
All materials in this class could enable room-temperature-scale moiré effects, a feature previously envisioned for BN~\cite{xian2019multiflat,dang2025twisting,PhysRevB.108.075109,roman2023excitons,apelian2024delocalization,yananose2025metamorphic}. 
Unlike BN, however, several exhibit much smaller monolayer gaps -- for instance, GaP’s gap of $1.55$eV is comparable to some TMDs and four times smaller than BN -- making electrostatic doping feasible and enabling experimental access to this room-temperature moiré physics.

\begin{figure}
\centering
\includegraphics[width=\columnwidth]{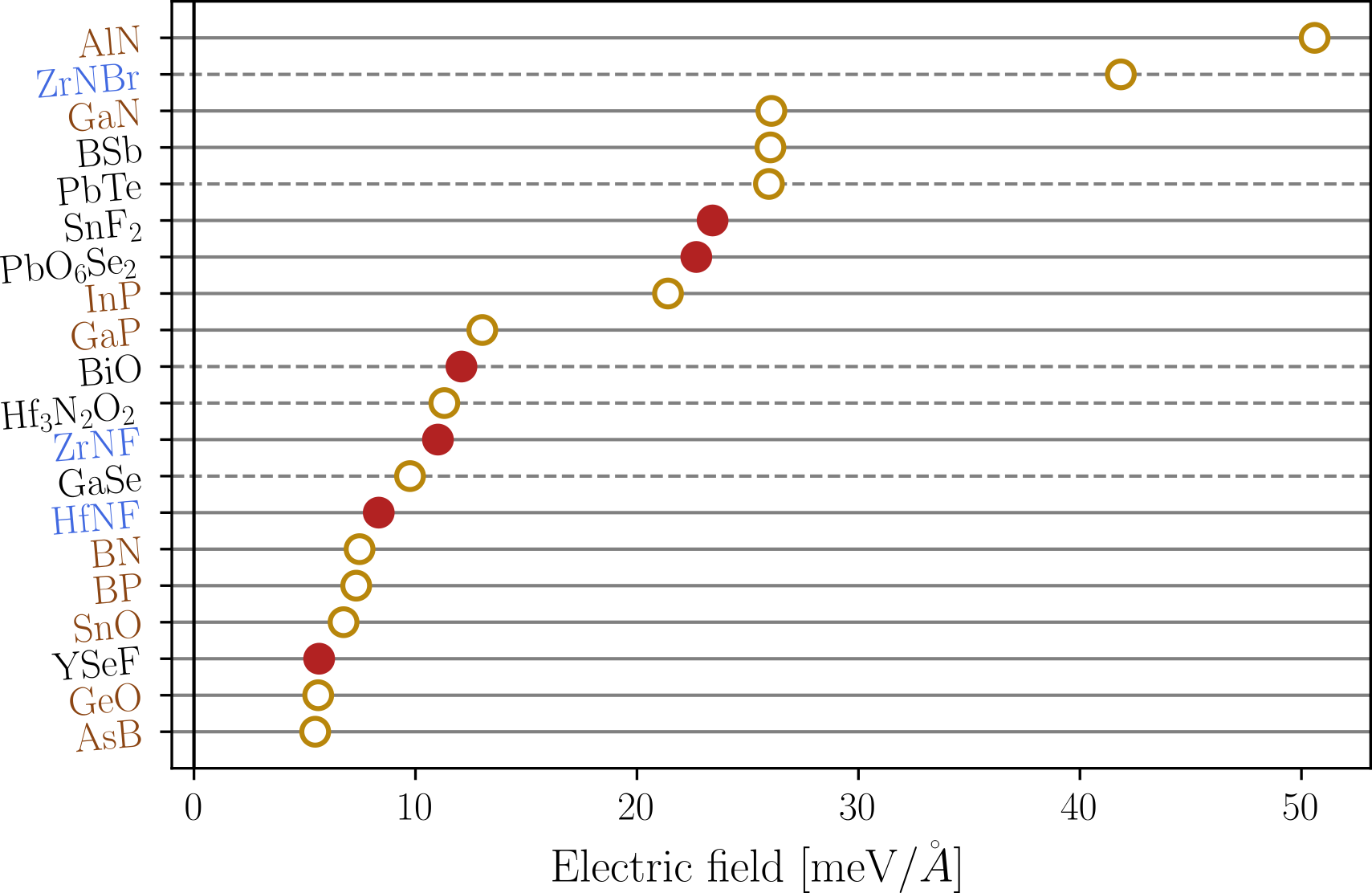}
\caption{\textit{
Electric field amplitude produced within the ferroelectric domains for the 20 $K$-valley semiconductors with the highest values. Symbols, colors, and linestyles follow those of Fig.~\ref{fig_fullclassification}. 
The spin-orbit coupled listed ZrNBr in the second row has a different lattice structure but identical chemical composition than the ZrNBr listed in Fig.~\ref{fig_fullclassification} without spin-orbit coupling; see full list of parameters in Methods.
}}
\label{fig_polar_field}
\end{figure}

Even if these compounds are not electrostatically dopable, they can exhibit large spatially varying polar/ferroelectric domains that are tunable by twist, as previously found in twisted BN~\cite{kim2024electrostatic}. From our DFT calculations, we extract the amplitude of the electric field generated by these domains and identify the materials producing the strongest fields. Fig~\ref{fig_polar_field} shows the electric field amplitude for the 20 highest-ranked materials, where the binary compounds highlighted above occupy most of the top positions. Such potentials may be used as a template to engineer flat bands in other 2d materials~\cite{ghorashi2023topological,ghorashi2023multilayer,zeng2024gate,seleznev2024inducing,ault2025optimizing,guerci2022designer}, or design electrically tunable moiré-domains~\cite{yasuda2021stacking} and phonons~\cite{ramos2025flat}.

The family with the next largest band gaps is the transition metal phosphorus trichalcogenides (TMPT, either MPS$_3$ or MM'P$_2$S$_6$ with M $\in \{ {\rm Sn}, {\rm Pb} , {\rm Zn}, {\rm Mg}, \cdots  \}$ and X $\in \{ {\rm S}, {\rm Se} \}$, green labels in Fig.~\ref{fig_fullclassification}). This family exhibits robust gaps of $\sim 10-20$meV across all investigated compounds in the conduction band. Further, their magic angle lies safely above the relaxation angle, ensuring that our high-throughput estimates are reliable for these materials and providing a clear experimental guide for sample design.

\section{Acknowledgements}

We thank A. Tyner, D. Kaplan, and J. Pixley for a discussion on related topics. 
VC also acknowledges helpful conversations with N. Regnault, C. de Beule and M. Albert. 
J.C. acknowledges support from the Air Force Office of Scientific Research under Grant No. FA9550-24-1-0222. 
N.N. acknowledges the support from the JSPS Overseas Research Fellowship.  
The Flatiron Institute is a division of the Simons Foundation.

\section{Data availability}

The data required to reproduce the results presented in this manuscript are compiled in the parameter tables provided in the Methods section. Additionally, all bilayer parameters and the relaxed structure files from the commensurate DFT calculations are available in this \href{https://doi.org/10.5281/zenodo.17903215}{Zenodo repository}.

\bibliography{biblio}

@article{geim2013van,
  title={Van der Waals heterostructures},
  author={Geim, Andre K and Grigorieva, Irina V},
  journal={Nature},
  volume={499},
  number={7459},
  pages={419--425},
  year={2013},
  publisher={Nature Publishing Group UK London}
}

@article{liu2016van,
  title={Van der Waals heterostructures and devices},
  author={Liu, Yuan and Weiss, Nathan O and Duan, Xidong and Cheng, Hung-Chieh and Huang, Yu and Duan, Xiangfeng},
  journal={Nature Reviews Materials},
  volume={1},
  number={9},
  pages={1--17},
  year={2016},
  publisher={Nature Publishing Group}
}

@article{novoselov20162d,
  title={2D materials and van der Waals heterostructures},
  author={Novoselov, K Sꎬ and Mishchenko, Artem and Carvalho, Alexandra and Castro Neto, AH},
  journal={Science},
  volume={353},
  number={6298},
  pages={aac9439},
  year={2016},
  publisher={American Association for the Advancement of Science}
}

@article{castellanos2022van,
  title={Van der Waals heterostructures},
  author={Castellanos-Gomez, Andres and Duan, Xiangfeng and Fei, Zhe and Gutierrez, Humberto Rodriguez and Huang, Yuan and Huang, Xinyu and Quereda, Jorge and Qian, Qi and Sutter, Eli and Sutter, Peter},
  journal={Nature Reviews Methods Primers},
  volume={2},
  number={1},
  pages={58},
  year={2022},
  publisher={Nature Publishing Group UK London}
}

@article{cao2018unconventional,
  title={Unconventional superconductivity in magic-angle graphene superlattices},
  author={Cao, Yuan and Fatemi, Valla and Fang, Shiang and Watanabe, Kenji and Taniguchi, Takashi and Kaxiras, Efthimios and Jarillo-Herrero, Pablo},
  journal={Nature},
  volume={556},
  number={7699},
  pages={43--50},
  year={2018},
  publisher={Nature Publishing Group UK London}
}

@article{yankowitz2019tuning,
  title={Tuning superconductivity in twisted bilayer graphene},
  author={Yankowitz, Matthew and Chen, Shaowen and Polshyn, Hryhoriy and Zhang, Yuxuan and Watanabe, Kenji and Taniguchi, Takashi and Graf, David and Young, Andrea F and Dean, Cory R},
  journal={Science},
  volume={363},
  number={6431},
  pages={1059--1064},
  year={2019},
  publisher={American Association for the Advancement of Science}
}

@article{andrei2021marvels,
  title={The marvels of moir{\'e} materials},
  author={Andrei, Eva Y and Efetov, Dmitri K and Jarillo-Herrero, Pablo and MacDonald, Allan H and Mak, Kin Fai and Senthil, T and Tutuc, Emanuel and Yazdani, Ali and Young, Andrea F},
  journal={Nature Reviews Materials},
  volume={6},
  number={3},
  pages={201--206},
  year={2021},
  publisher={Nature Publishing Group UK London}
}

@article{mak2022semiconductor,
  title={Semiconductor moir{\'e} materials},
  author={Mak, Kin Fai and Shan, Jie},
  journal={Nature Nanotechnology},
  volume={17},
  number={7},
  pages={686--695},
  year={2022},
  publisher={Nature Publishing Group UK London}
}

@article{xiao2020moire,
  title={Moir{\'e} is more: access to new properties of two-dimensional layered materials},
  author={Xiao, Yao and Liu, Jinglu and Fu, Lei},
  journal={Matter},
  volume={3},
  number={4},
  pages={1142--1161},
  year={2020},
  publisher={Elsevier}
}

@article{kennes2021moire,
  title={Moir{\'e} heterostructures as a condensed-matter quantum simulator},
  author={Kennes, Dante M and Claassen, Martin and Xian, Lede and Georges, Antoine and Millis, Andrew J and Hone, James and Dean, Cory R and Basov, DN and Pasupathy, Abhay N and Rubio, Angel},
  journal={Nature Physics},
  volume={17},
  number={2},
  pages={155--163},
  year={2021},
  publisher={Nature Publishing Group UK London}
}

@article{redekop2024direct,
  title={Direct magnetic imaging of fractional Chern insulators in twisted MoTe2},
  author={Redekop, Evgeny and Zhang, Canxun and Park, Heonjoon and Cai, Jiaqi and Anderson, Eric and Sheekey, Owen and Arp, Trevor and Babikyan, Grigory and Salters, Samuel and Watanabe, Kenji and others},
  journal={Nature},
  volume={635},
  number={8039},
  pages={584--589},
  year={2024},
  publisher={Nature Publishing Group UK London}
}

@article{lu2019superconductors,
  title={Superconductors, orbital magnets and correlated states in magic-angle bilayer graphene},
  author={Lu, Xiaobo and Stepanov, Petr and Yang, Wei and Xie, Ming and Aamir, Mohammed Ali and Das, Ipsita and Urgell, Carles and Watanabe, Kenji and Taniguchi, Takashi and Zhang, Guangyu and others},
  journal={Nature},
  volume={574},
  number={7780},
  pages={653--657},
  year={2019},
  publisher={Nature Publishing Group UK London}
}

@article{jiang20242d,
  title={2D theoretically twistable material database},
  author={Jiang, Yi and Petralanda, Urko and Skorupskii, Grigorii and Xu, Qiaoling and Pi, Hanqi and C{\u{a}}lug{\u{a}}ru, Dumitru and Hu, Haoyu and Xie, Jiaze and Mustaf, Rose Albu and H{\"o}hn, Peter and others},
  journal={arXiv preprint arXiv:2411.09741},
  year={2024}
}

@article{jung2014ab,
  title={Ab initio theory of moir{\'e} superlattice bands in layered two-dimensional materials},
  author={Jung, Jeil and Raoux, Arnaud and Qiao, Zhenhua and MacDonald, Allan H},
  journal={Physical Review B},
  volume={89},
  number={20},
  pages={205414},
  year={2014},
  publisher={APS}
}

@article{koshino2020effective,
  title={Effective continuum model for relaxed twisted bilayer graphene and moir{\'e} electron-phonon interaction},
  author={Koshino, Mikito and Nam, Nguyen NT},
  journal={Physical Review B},
  volume={101},
  number={19},
  pages={195425},
  year={2020},
  publisher={APS}
}

@article{zhang2025twist,
  title={Twist-angle transferable continuum model and second flat Chern band in twisted MoTe2 and WSe2},
  author={Zhang, Xiao-Wei and Yang, Kaijie and Wang, Chong and Liu, Xiaoyu and Cao, Ting and Xiao, Di},
  journal={npj Quantum Materials},
  year={2025},
  publisher={Nature Publishing Group UK London}
}

@article{chang2013experimental,
  title={Experimental observation of the quantum anomalous Hall effect in a magnetic topological insulator},
  author={Chang, Cui-Zu and Zhang, Jinsong and Feng, Xiao and Shen, Jie and Zhang, Zuocheng and Guo, Minghua and Li, Kang and Ou, Yunbo and Wei, Pang and Wang, Li-Li and others},
  journal={Science},
  volume={340},
  number={6129},
  pages={167--170},
  year={2013},
  publisher={American Association for the Advancement of Science}
}

@article{benschop2021measuring,
  title={Measuring local moir{\'e} lattice heterogeneity of twisted bilayer graphene},
  author={Benschop, Tjerk and de Jong, Tobias A and Stepanov, Petr and Lu, Xiaobo and Stalman, Vincent and van der Molen, Sense Jan and Efetov, Dmitri K and Allan, Milan P},
  journal={Physical Review Research},
  volume={3},
  number={1},
  pages={013153},
  year={2021},
  publisher={APS}
}

@article{kapfer2023programming,
  title={Programming twist angle and strain profiles in 2D materials},
  author={Kapfer, Ma{\"e}lle and Jessen, Bjarke S and Eisele, Megan E and Fu, Matthew and Danielsen, Dorte R and Darlington, Thomas P and Moore, Samuel L and Finney, Nathan R and Marchese, Ariane and Hsieh, Valerie and others},
  journal={Science},
  volume={381},
  number={6658},
  pages={677--681},
  year={2023},
  publisher={American Association for the Advancement of Science}
}

@article{lai2025moire,
  title={Moir{\'e} periodic and quasiperiodic crystals in heterostructures of twisted bilayer graphene on hexagonal boron nitride},
  author={Lai, Xinyuan and Li, Guohong and Coe, Angela M and Pixley, Jedediah H and Watanabe, Kenji and Taniguchi, Takashi and Andrei, Eva Y},
  journal={Nature materials},
  pages={1--8},
  year={2025},
  publisher={Nature Publishing Group UK London}
}

@article{petralanda2024two,
  title={Two-dimensional topological quantum chemistry and catalog of topological materials},
  author={Petralanda, Urko and Jiang, Yi and Bernevig, B Andrei and Regnault, Nicolas and Elcoro, Luis},
  journal={arXiv preprint arXiv:2411.08950},
  year={2024}
}

@article{wang2020correlated,
  title={Correlated electronic phases in twisted bilayer transition metal dichalcogenides},
  author={Wang, Lei and Shih, En-Min and Ghiotto, Augusto and Xian, Lede and Rhodes, Daniel A and Tan, Cheng and Claassen, Martin and Kennes, Dante M and Bai, Yusong and Kim, Bumho and others},
  journal={Nature materials},
  volume={19},
  number={8},
  pages={861--866},
  year={2020},
  publisher={Nature Publishing Group UK London}
}

@article{guo2025superconductivity,
  title={Superconductivity in 5.0° twisted bilayer WSe2},
  author={Guo, Yinjie and Pack, Jordan and Swann, Joshua and Holtzman, Luke and Cothrine, Matthew and Watanabe, Kenji and Taniguchi, Takashi and Mandrus, David G and Barmak, Katayun and Hone, James and others},
  journal={Nature},
  volume={637},
  number={8047},
  pages={839--845},
  year={2025},
  publisher={Nature Publishing Group UK London}
}

@article{xia2025superconductivity,
  title={Superconductivity in twisted bilayer WSe2},
  author={Xia, Yiyu and Han, Zhongdong and Watanabe, Kenji and Taniguchi, Takashi and Shan, Jie and Mak, Kin Fai},
  journal={Nature},
  volume={637},
  number={8047},
  pages={833--838},
  year={2025},
  publisher={Nature Publishing Group UK London}
}

@article{cai2023signatures,
  title={Signatures of fractional quantum anomalous Hall states in twisted MoTe2},
  author={Cai, Jiaqi and Anderson, Eric and Wang, Chong and Zhang, Xiaowei and Liu, Xiaoyu and Holtzmann, William and Zhang, Yinong and Fan, Fengren and Taniguchi, Takashi and Watanabe, Kenji and others},
  journal={Nature},
  volume={622},
  number={7981},
  pages={63--68},
  year={2023},
  publisher={Nature Publishing Group UK London}
}

@article{xu2023observation,
  title={Observation of integer and fractional quantum anomalous Hall effects in twisted bilayer MoTe 2},
  author={Xu, Fan and Sun, Zheng and Jia, Tongtong and Liu, Chang and Xu, Cheng and Li, Chushan and Gu, Yu and Watanabe, Kenji and Taniguchi, Takashi and Tong, Bingbing and others},
  journal={Physical Review X},
  volume={13},
  number={3},
  pages={031037},
  year={2023},
  publisher={APS}
}

@article{kang2024evidence,
  title={Evidence of the fractional quantum spin Hall effect in moir{\'e} MoTe2},
  author={Kang, Kaifei and Shen, Bowen and Qiu, Yichen and Zeng, Yihang and Xia, Zhengchao and Watanabe, Kenji and Taniguchi, Takashi and Shan, Jie and Mak, Kin Fai},
  journal={Nature},
  volume={628},
  number={8008},
  pages={522--526},
  year={2024},
  publisher={Nature Publishing Group UK London}
}

@article{massatt2017electronic,
  title={Electronic density of states for incommensurate layers},
  author={Massatt, Daniel and Luskin, Mitchell and Ortner, Christoph},
  journal={Multiscale Modeling \& Simulation},
  volume={15},
  number={1},
  pages={476--499},
  year={2017},
  publisher={SIAM}
}

@article{park2025observation,
  title={Observation of High-Temperature Dissipationless Fractional Chern Insulator},
  author={Park, Heonjoon and Li, Weijie and Hu, Chaowei and Beach, Christiano and Gon{\c{c}}alves, Miguel and Mendez-Valderrama, Juan Felipe and Herzog-Arbeitman, Jonah and Taniguchi, Takashi and Watanabe, Kenji and Cobden, David and others},
  journal={arXiv preprint arXiv:2503.10989},
  year={2025}
}

@article{ezzi2024analytical,
  title={Analytical model for atomic relaxation in twisted moir{\'e} materials},
  author={Ezzi, Mohammed M Al and Pallewela, Gayani N and De Beule, Christophe and Mele, EJ and Adam, Shaffique},
  journal={Physical Review Letters},
  volume={133},
  number={26},
  pages={266201},
  year={2024},
  publisher={APS}
}

@article{wang2024fractional,
  title={Fractional Chern insulator in twisted bilayer MoTe 2},
  author={Wang, Chong and Zhang, Xiao-Wei and Liu, Xiaoyu and He, Yuchi and Xu, Xiaodong and Ran, Ying and Cao, Ting and Xiao, Di},
  journal={Physical Review Letters},
  volume={132},
  number={3},
  pages={036501},
  year={2024},
  publisher={APS}
}

@article{liu2025symmetry,
  title={Symmetry-enforced Moir$\backslash$'e Topology},
  author={Liu, Yunzhe and Yang, Kaijie and Liu, Chao-Xing and Yu, Jiabin},
  journal={arXiv preprint arXiv:2509.06906},
  year={2025}
}

@article{yang2024topological,
  title={Topological minibands and interaction driven quantum anomalous Hall state in topological insulator based moir{\'e} heterostructures},
  author={Yang, Kaijie and Xu, Zian and Feng, Yanjie and Schindler, Frank and Xu, Yuanfeng and Bi, Zhen and Bernevig, B Andrei and Tang, Peizhe and Liu, Chao-Xing},
  journal={Nature communications},
  volume={15},
  number={1},
  pages={2670},
  year={2024},
  publisher={Nature Publishing Group UK London}
}

@article{choi2025superconductivity,
  title={Superconductivity and quantized anomalous Hall effect in rhombohedral graphene},
  author={Choi, Youngjoon and Choi, Ysun and Valentini, Marco and Patterson, Caitlin L and Holleis, Ludwig FW and Sheekey, Owen I and Stoyanov, Hari and Cheng, Xiang and Taniguchi, Takashi and Watanabe, Kenji and others},
  journal={Nature},
  pages={1--6},
  year={2025},
  publisher={Nature Publishing Group UK London}
}

@article{zhou2021superconductivity,
  title={Superconductivity in rhombohedral trilayer graphene},
  author={Zhou, Haoxin and Xie, Tian and Taniguchi, Takashi and Watanabe, Kenji and Young, Andrea F},
  journal={Nature},
  volume={598},
  number={7881},
  pages={434--438},
  year={2021},
  publisher={Nature Publishing Group UK London}
}

@article{han2025signatures,
  title={Signatures of chiral superconductivity in rhombohedral graphene},
  author={Han, Tonghang and Lu, Zhengguang and Hadjri, Zach and Shi, Lihan and Wu, Zhenghan and Xu, Wei and Yao, Yuxuan and Cotten, Armel A and Sharifi Sedeh, Omid and Weldeyesus, Henok and others},
  journal={Nature},
  volume={643},
  number={8072},
  pages={654--661},
  year={2025},
  publisher={Nature Publishing Group UK London}
}

@article{zhang2024polarization,
  title={Polarization-driven band topology evolution in twisted MoTe2 and WSe2},
  author={Zhang, Xiao-Wei and Wang, Chong and Liu, Xiaoyu and Fan, Yueyao and Cao, Ting and Xiao, Di},
  journal={Nature Communications},
  volume={15},
  number={1},
  pages={4223},
  year={2024},
  publisher={Nature Publishing Group UK London}
}

@article{enaldiev2020stacking,
  title={Stacking domains and dislocation networks in marginally twisted bilayers of transition metal dichalcogenides},
  author={Enaldiev, VV and Z{\'o}lyomi, Viktor and Yelgel, CELAL and Magorrian, SJ and Fal’Ko, VI},
  journal={Physical review letters},
  volume={124},
  number={20},
  pages={206101},
  year={2020},
  publisher={APS}
}

@article{nakatsuji2025moir,
  title={Moir$\backslash$'e Band Engineering in Twisted Trilayer WSe2},
  author={Nakatsuji, Naoto and Kawakami, Takuto and Tateishi, Hayato and Kato, Koichiro and Koshino, Mikito},
  journal={arXiv preprint arXiv:2504.20449},
  year={2025}
}

@article{nakatsuji2023multiscale,
  title={Multiscale lattice relaxation in general twisted trilayer graphenes},
  author={Nakatsuji, Naoto and Kawakami, Takuto and Koshino, Mikito},
  journal={Physical Review X},
  volume={13},
  number={4},
  pages={041007},
  year={2023},
  publisher={APS}
}

@article{suzuura2002phonons,
  title={Phonons and electron-phonon scattering in carbon nanotubes},
  author={Suzuura, Hidekatsu and Ando, Tsuneya},
  journal={parameters},
  volume={10},
  number={2},
  pages={2},
  year={2002}
}

@article{hoke2024imaging,
  title={Imaging supermoire relaxation and conductive domain walls in helical trilayer graphene},
  author={Hoke, Jesse C and Li, Yifan and Hu, Yuwen and May-Mann, Julian and Watanabe, Kenji and Taniguchi, Takashi and Devakul, Trithep and Feldman, Benjamin E},
  journal={arXiv preprint arXiv:2410.16269},
  year={2024}
}

@article{zhou2023non,
  title={Non-Abelian topological superconductivity in maximally twisted double-layer spin-triplet valley-singlet superconductors},
  author={Zhou, Benjamin T and Egan, Shannon and Kush, Dhruv and Franz, Marcel},
  journal={Communications Physics},
  volume={6},
  number={1},
  pages={47},
  year={2023},
  publisher={Nature Publishing Group UK London}
}

@article{zhu2025wavefunction,
  title={Wavefunction textures in twisted bilayer graphene from first principles},
  author={Zhu, Albert and Bennett, Daniel and Larson, Daniel T and Ezzi, Mohammed M Al and Manousakis, Efstratios and Kaxiras, Efthimios},
  journal={arXiv preprint arXiv:2507.03675},
  year={2025}
}

@article{yananose2021chirality,
  title={Chirality-induced spin texture switching in twisted bilayer graphene},
  author={Yananose, Kunihiro and Cantele, Giovanni and Lucignano, Procolo and Cheong, Sang-Wook and Yu, Jaejun and Stroppa, Alessandro},
  journal={Physical Review B},
  volume={104},
  number={7},
  pages={075407},
  year={2021},
  publisher={APS}
}

@article{lucignano2019crucial,
  title={Crucial role of atomic corrugation on the flat bands and energy gaps of twisted bilayer graphene at the magic angle 1.08deg},
  author={Lucignano, Procolo and Alf{\`e}, Dario and Cataudella, Vittorio and Ninno, Domenico and Cantele, Giovanni},
  journal={Physical Review B},
  volume={99},
  number={19},
  pages={195419},
  year={2019},
  publisher={APS}
}

@article{shaidu2025transferable,
  title={Transferable dispersion-aware machine learning interatomic potentials for multilayer transition metal dichalcogenide heterostructures},
  author={Shaidu, Yusuf and Naik, Mit H and Louie, Steven G and Neaton, Jeffrey B},
  journal={npj Computational Materials},
  volume={11},
  number={1},
  pages={273},
  year={2025},
  publisher={Nature Publishing Group UK London}
}

@article{cano2021band,
  title={Band representations and topological quantum chemistry},
  author={Cano, Jennifer and Bradlyn, Barry},
  journal={Annual Review of Condensed Matter Physics},
  volume={12},
  number={1},
  pages={225--246},
  year={2021},
  publisher={Annual Reviews}
}

@article{crepel2025topologically,
  title={Topologically protected flatness in chiral moir{\'e} heterostructures},
  author={Cr{\'e}pel, Valentin and Ding, Peize and Verma, Nishchhal and Regnault, Nicolas and Queiroz, Raquel},
  journal={Physical Review X},
  volume={15},
  number={2},
  pages={021056},
  year={2025},
  publisher={APS}
}

@article{crepel2023chiral,
  title={Chiral model of twisted bilayer graphene realized in a monolayer},
  author={Cr{\'e}pel, Valentin and Dunbrack, Aaron and Guerci, Daniele and Bonini, John and Cano, Jennifer},
  journal={Physical Review B},
  volume={108},
  number={7},
  pages={075126},
  year={2023},
  publisher={APS}
}

@article{jia2024moire,
  title={Moir{\'e} fractional Chern insulators. I. First-principles calculations and continuum models of twisted bilayer MoTe 2},
  author={Jia, Yujin and Yu, Jiabin and Liu, Jiaxuan and Herzog-Arbeitman, Jonah and Qi, Ziyue and Pi, Hanqi and Regnault, Nicolas and Weng, Hongming and Bernevig, B Andrei and Wu, Quansheng},
  journal={Physical Review B},
  volume={109},
  number={20},
  pages={205121},
  year={2024},
  publisher={APS}
}

@article{tilak2022moire,
  title={Moir{\'e} potential, lattice relaxation, and layer polarization in marginally twisted MoS2 bilayers},
  author={Tilak, Nikhil and Li, Guohong and Taniguchi, Takashi and Watanabe, Kenji and Andrei, Eva Y},
  journal={Nano Letters},
  volume={23},
  number={1},
  pages={73--81},
  year={2022},
  publisher={ACS Publications}
}

@article{li2021van,
  title={Van der Waals epitaxial growth of air-stable CrSe2 nanosheets with thickness-tunable magnetic order},
  author={Li, Bo and Wan, Zhong and Wang, Cong and Chen, Peng and Huang, Bevin and Cheng, Xing and Qian, Qi and Li, Jia and Zhang, Zhengwei and Sun, Guangzhuang and others},
  journal={Nature materials},
  volume={20},
  number={6},
  pages={818--825},
  year={2021},
  publisher={Nature Publishing Group UK London}
}

@article{van1980crse2,
  title={CrSe2, a new layered dichalcogenide},
  author={Van Bruggen, CF and Haange, RJ and Wiegers, GA and De Boer, DKG},
  journal={Physica B+ c},
  volume={99},
  number={1-4},
  pages={166--172},
  year={1980},
  publisher={Elsevier}
}

@article{wu2022plane,
  title={In-plane epitaxy-strain-tuning intralayer and interlayer magnetic coupling in CrSe 2 and CrTe 2 monolayers and bilayers},
  author={Wu, Linlu and Zhou, Linwei and Zhou, Xieyu and Wang, Cong and Ji, Wei},
  journal={Physical Review B},
  volume={106},
  number={8},
  pages={L081401},
  year={2022},
  publisher={APS}
}

@article{xiao2022novel,
  title={Novel two-dimensional ferromagnetic materials CrX 2 (X= O, S, Se) with high Curie temperature},
  author={Xiao, Gang and Xiao, Wen-Zhi and Chen, Qiao and Wang, Ling-ling},
  journal={Journal of Materials Chemistry C},
  volume={10},
  number={46},
  pages={17665--17674},
  year={2022},
  publisher={Royal Society of Chemistry}
}

@article{fan2024progress,
  title={Progress in the preparation and physical properties of two-dimensional Cr-based chalcogenide materials and heterojunctions},
  author={Fan, Xiulian and Xin, Ruifeng and Li, Li and Zhang, Bo and Li, Cheng and Zhou, Xilong and Chen, Huanzhi and Zhang, Hongyan and OuYang, Fangping and Zhou, Yu},
  journal={Frontiers of Physics},
  volume={19},
  number={2},
  pages={23401},
  year={2024},
  publisher={Springer}
}

@article{gonccalves2025spinless,
  title={Spinless and spinful charge excitations in moir\'e Fractional Chern Insulators},
  author={Gon{\c{c}}alves, Miguel and Mendez-Valderrama, Juan Felipe and Herzog-Arbeitman, Jonah and Yu, Jiabin and Xu, Xiaodong and Xiao, Di and Bernevig, B Andrei and Regnault, Nicolas},
  journal={arXiv preprint arXiv:2506.05330},
  year={2025}
}

@article{zhang2021room,
  title={Room-temperature intrinsic ferromagnetism in epitaxial CrTe2 ultrathin films},
  author={Zhang, Xiaoqian and Lu, Qiangsheng and Liu, Wenqing and Niu, Wei and Sun, Jiabao and Cook, Jacob and Vaninger, Mitchel and Miceli, Paul F and Singh, David J and Lian, Shang-Wei and others},
  journal={Nature communications},
  volume={12},
  number={1},
  pages={2492},
  year={2021},
  publisher={Nature Publishing Group UK London}
}

@article{xian2022spin,
  title={Spin mapping of intralayer antiferromagnetism and field-induced spin reorientation in monolayer CrTe2},
  author={Xian, Jing-Jing and Wang, Cong and Nie, Jin-Hua and Li, Rui and Han, Mengjiao and Lin, Junhao and Zhang, Wen-Hao and Liu, Zhen-Yu and Zhang, Zhi-Mo and Miao, Mao-Peng and others},
  journal={Nature communications},
  volume={13},
  number={1},
  pages={257},
  year={2022},
  publisher={Nature Publishing Group UK London}
}

@article{nuckolls2024microscopic,
  title={A microscopic perspective on moir{\'e} materials},
  author={Nuckolls, Kevin P and Yazdani, Ali},
  journal={Nature Reviews Materials},
  volume={9},
  number={7},
  pages={460--480},
  year={2024},
  publisher={Nature Publishing Group UK London}
}

@article{hasan2025beyond,
  title={beyond the surface—exploring the complexities of 2D materials with density functional theory},
  author={Hasan, Mohammed Abdulabbas and Abdulhussein, Heider A},
  journal={Journal of Materials Science},
  pages={1--45},
  year={2025},
  publisher={Springer}
}

@article{chen2023moire,
  title={Moir{\'e} heterostructures: highly tunable platforms for quantum simulation and future computing},
  author={Chen, Moyu and Chen, Fanqiang and Cheng, Bin and Liang, Shi Jun and Miao, Feng},
  journal={Journal of Semiconductors},
  volume={44},
  number={1},
  pages={010301--1},
  year={2023},
  publisher={Journal of Semiconductors}
}

@article{zhang2022n,
  title={O (N) ab initio calculation scheme for large-scale moir{\'e} structures},
  author={Zhang, Tan and Regnault, Nicolas and Bernevig, B Andrei and Dai, Xi and Weng, Hongming},
  journal={Physical Review B},
  volume={105},
  number={12},
  pages={125127},
  year={2022},
  publisher={APS}
}

@article{verma2020status,
  title={Status and challenges of density functional theory},
  author={Verma, Pragya and Truhlar, Donald G},
  journal={Trends in Chemistry},
  volume={2},
  number={4},
  pages={302--318},
  year={2020},
  publisher={Elsevier}
}

@article{carr2020electronic,
  title={Electronic-structure methods for twisted moir{\'e} layers},
  author={Carr, Stephen and Fang, Shiang and Kaxiras, Efthimios},
  journal={Nature Reviews Materials},
  volume={5},
  number={10},
  pages={748--763},
  year={2020},
  publisher={Nature Publishing Group UK London}
}

@article{de2025theory,
  title={Theory for Lattice Relaxation in Marginal Twist Moires},
  author={De Beule, Christophe and Pallewela, Gayani N and Ezzi, Mohammed M Al and Peng, Liangtao and Mele, EJ and Adam, Shaffique},
  journal={arXiv preprint arXiv:2503.19162},
  year={2025}
}

@article{lei2025moire,
  title={Moir{\'e} band theory for M-valley twisted transition metal dichalcogenides},
  author={Lei, Chao and Mahon, Perry T and MacDonald, Allan H},
  journal={Physical Review Letters},
  volume={135},
  number={19},
  pages={196402},
  year={2025},
  publisher={APS}
}

@article{naik2022twister,
  title={Twister: Construction and structural relaxation of commensurate moir{\'e} superlattices},
  author={Naik, Saismit and Naik, Mit H and Maity, Indrajit and Jain, Manish},
  journal={Computer Physics Communications},
  volume={271},
  pages={108184},
  year={2022},
  publisher={Elsevier}
}

@article{mcgilly2020visualization,
  title={Visualization of moir{\'e} superlattices},
  author={McGilly, Leo J and Kerelsky, Alexander and Finney, Nathan R and Shapovalov, Konstantin and Shih, En-Min and Ghiotto, Augusto and Zeng, Yihang and Moore, Samuel L and Wu, Wenjing and Bai, Yusong and others},
  journal={Nature Nanotechnology},
  volume={15},
  number={7},
  pages={580--584},
  year={2020},
  publisher={Nature Publishing Group UK London}
}

@article{cazeaux2023relaxation,
  title={Relaxation and domain wall structure of bilayer moir{\'e} systems},
  author={Cazeaux, Paul and Clark, Drake and Engelke, Rebecca and Kim, Philip and Luskin, Mitchell},
  journal={Journal of Elasticity},
  volume={154},
  number={1},
  pages={443--466},
  year={2023},
  publisher={Springer}
}

@article{chen2025quantum,
  title={Quantum-geometric dipole: a topological boost to flavor ferromagnetism in flat bands},
  author={Chen, Lei and Ghorashi, Sayed Ali Akbar and Cano, Jennifer and Cr{\'e}pel, Valentin},
  journal={arXiv preprint arXiv:2506.22417},
  year={2025}
}

@article{bistritzer2011moire,
  title={Moir{\'e} bands in twisted double-layer graphene},
  author={Bistritzer, Rafi and MacDonald, Allan H},
  journal={Proceedings of the National Academy of Sciences},
  volume={108},
  number={30},
  pages={12233--12237},
  year={2011},
  publisher={National Academy of Sciences}
}

@article{carr2018relaxation,
  title={Relaxation and domain formation in incommensurate two-dimensional heterostructures},
  author={Carr, Stephen and Massatt, Daniel and Torrisi, Steven B and Cazeaux, Paul and Luskin, Mitchell and Kaxiras, Efthimios},
  journal={Physical Review B},
  volume={98},
  number={22},
  pages={224102},
  year={2018},
  publisher={APS}
}

@article{po2017symmetry,
  title={Symmetry-based indicators of band topology in the 230 space groups},
  author={Po, Hoi Chun and Vishwanath, Ashvin and Watanabe, Haruki},
  journal={Nature communications},
  volume={8},
  number={1},
  pages={50},
  year={2017},
  publisher={Nature Publishing Group UK London}
}

@article{slater1952soluble,
  title={A soluble problem in energy bands},
  author={Slater, JC},
  journal={Physical Review},
  volume={87},
  number={5},
  pages={807},
  year={1952},
  publisher={APS}
}

@article{wu2019topological,
  title={Topological insulators in twisted transition metal dichalcogenide homobilayers},
  author={Wu, Fengcheng and Lovorn, Timothy and Tutuc, Emanuel and Martin, Ivar and MacDonald, AH},
  journal={Physical review letters},
  volume={122},
  number={8},
  pages={086402},
  year={2019},
  publisher={APS}
}

@article{bradlyn2017topological,
  title={Topological quantum chemistry},
  author={Bradlyn, Barry and Elcoro, Luis and Cano, Jennifer and Vergniory, Maia G and Wang, Zhijun and Felser, Claudia and Aroyo, Mois I and Bernevig, B Andrei},
  journal={Nature},
  volume={547},
  number={7663},
  pages={298--305},
  year={2017},
  publisher={Nature Publishing Group UK London}
}

@article{fang2012bulk,
  title={Bulk topological invariants in noninteracting point group symmetric insulators},
  author={Fang, Chen and Gilbert, Matthew J and Bernevig, B Andrei},
  journal={Physical Review B—Condensed Matter and Materials Physics},
  volume={86},
  number={11},
  pages={115112},
  year={2012},
  publisher={APS}
}

@article{devakul2021magic,
  title={Magic in twisted transition metal dichalcogenide bilayers},
  author={Devakul, Trithep and Cr{\'e}pel, Valentin and Zhang, Yang and Fu, Liang},
  journal={Nature communications},
  volume={12},
  number={1},
  pages={6730},
  year={2021},
  publisher={Nature Publishing Group UK London}
}

@article{crepel2025efficient,
  title={Efficient prediction of superlattice and anomalous miniband topology from quantum geometry},
  author={Cr{\'e}pel, Valentin and Cano, Jennifer},
  journal={Physical Review X},
  volume={15},
  number={1},
  pages={011004},
  year={2025},
  publisher={APS}
}

@article{crepel2023anomalous,
  title={Anomalous Hall metal and fractional Chern insulator in twisted transition metal dichalcogenides},
  author={Cr{\'e}pel, Valentin and Fu, Liang},
  journal={Physical Review B},
  volume={107},
  number={20},
  pages={L201109},
  year={2023},
  publisher={APS}
}

@article{anderson2023programming,
  title={Programming correlated magnetic states with gate-controlled moir{\'e} geometry},
  author={Anderson, Eric and Fan, Feng-Ren and Cai, Jiaqi and Holtzmann, William and Taniguchi, Takashi and Watanabe, Kenji and Xiao, Di and Yao, Wang and Xu, Xiaodong},
  journal={Science},
  volume={381},
  number={6655},
  pages={325--330},
  year={2023},
  publisher={American Association for the Advancement of Science}
}

@article{nakagawa2021gate,
  title={Gate-controlled BCS-BEC crossover in a two-dimensional superconductor},
  author={Nakagawa, Yuji and Kasahara, Yuichi and Nomoto, Takuya and Arita, Ryotaro and Nojima, Tsutomu and Iwasa, Yoshihiro},
  journal={Science},
  volume={372},
  number={6538},
  pages={190--195},
  year={2021},
  publisher={American Association for the Advancement of Science}
}

@article{lu2024fractional,
  title={Fractional quantum anomalous Hall effect in multilayer graphene},
  author={Lu, Zhengguang and Han, Tonghang and Yao, Yuxuan and Reddy, Aidan P and Yang, Jixiang and Seo, Junseok and Watanabe, Kenji and Taniguchi, Takashi and Fu, Liang and Ju, Long},
  journal={Nature},
  volume={626},
  number={8000},
  pages={759--764},
  year={2024},
  publisher={Nature Publishing Group UK London}
}

@article{xie2025tunable,
  title={Tunable fractional Chern insulators in rhombohedral graphene superlattices},
  author={Xie, Jian and Huo, Zihao and Lu, Xin and Feng, Zuo and Zhang, Zaizhe and Wang, Wenxuan and Yang, Qiu and Watanabe, Kenji and Taniguchi, Takashi and Liu, Kaihui and others},
  journal={Nature Materials},
  pages={1--7},
  year={2025},
  publisher={Nature Publishing Group UK London}
}

@article{fischer2024theory,
  title={Theory of intervalley-coherent AFM order and topological superconductivity in tWSe $ \_2$},
  author={Fischer, Ammon and Klebl, Lennart and Cr{\'e}pel, Valentin and Ryee, Siheon and Rubio, Angel and Xian, Lede and Wehling, Tim O and Georges, Antoine and Kennes, Dante M and Millis, Andrew J},
  journal={arXiv preprint arXiv:2412.14296},
  year={2024}
}

@article{munoz2025twist,
  title={Twist-angle evolution of the intervalley-coherent antiferromagnet in twisted WSe 2},
  author={Mu{\~n}oz-Segovia, Daniel and Cr{\'e}pel, Valentin and Queiroz, Raquel and Millis, Andrew J},
  journal={Physical Review B},
  volume={112},
  number={8},
  pages={085111},
  year={2025},
  publisher={APS}
}

@article{waters2024interplay,
  title={Interplay of electronic crystals with integer and fractional Chern insulators in moir$\backslash$'e pentalayer graphene},
  author={Waters, Dacen and Okounkova, Anna and Su, Ruiheng and Zhou, Boran and Yao, Jiang and Watanabe, Kenji and Taniguchi, Takashi and Xu, Xiaodong and Zhang, Ya-Hui and Folk, Joshua and others},
  journal={arXiv preprint arXiv:2408.10133},
  year={2024}
}

@article{hase1999electronic,
  title={Electronic structure of superconducting layered zirconium and hafnium nitride},
  author={Hase, Izumi and Nishihara, Yoshikazu},
  journal={Physical Review B},
  volume={60},
  number={3},
  pages={1573},
  year={1999},
  publisher={APS}
}

@article{felser1999electronic,
  title={Electronic structures and instabilities of ZrNCl and HfNCl: implications for superconductivity in the doped compounds},
  author={Felser, Claudia and Seshadri, Ram},
  journal={Journal of Materials Chemistry},
  volume={9},
  number={2},
  pages={459--464},
  year={1999},
  publisher={Royal Society of Chemistry}
}

@article{kasahara2015unconventional,
  title={Unconventional superconductivity in electron-doped layered metal nitride halides MNX (M= Ti, Zr, Hf; X= Cl, Br, I)},
  author={Kasahara, Yuichi and Kuroki, Kazuhiko and Yamanaka, Shoji and Taguchi, Yasujiro},
  journal={Physica C: Superconductivity and its Applications},
  volume={514},
  pages={354--367},
  year={2015},
  publisher={Elsevier}
}

@article{crepel2022spin,
  title={Spin-triplet superconductivity from excitonic effect in doped insulators},
  author={Cr{\'e}pel, Valentin and Fu, Liang},
  journal={Proceedings of the National Academy of Sciences},
  volume={119},
  number={13},
  pages={e2117735119},
  year={2022},
  publisher={National Academy of Sciences}
}

@article{lhachemi2025efficient,
  title={Efficient prediction of topological superlattice bands with spin-orbit coupling},
  author={Lhachemi, M Nabil Y and Cr{\'e}pel, Valentin and Cano, Jennifer},
  journal={arXiv preprint arXiv:2511.17483},
  year={2025}
}

@article{haastrup2018computational,
  title={The Computational 2D Materials Database: high-throughput modeling and discovery of atomically thin crystals},
  author={Haastrup, Sten and Strange, Mikkel and Pandey, Mohnish and Deilmann, Thorsten and Schmidt, Per S and Hinsche, Nicki F and Gjerding, Morten N and Torelli, Daniele and Larsen, Peter M and Riis-Jensen, Anders C and others},
  journal={2D Materials},
  volume={5},
  number={4},
  pages={042002},
  year={2018},
  publisher={IOP Publishing}
}

@article{mounet2018two,
  title={Two-dimensional materials from high-throughput computational exfoliation of experimentally known compounds},
  author={Mounet, Nicolas and Gibertini, Marco and Schwaller, Philippe and Campi, Davide and Merkys, Andrius and Marrazzo, Antimo and Sohier, Thibault and Castelli, Ivano Eligio and Cepellotti, Andrea and Pizzi, Giovanni and others},
  journal={Nature nanotechnology},
  volume={13},
  number={3},
  pages={246--252},
  year={2018},
  publisher={Nature Publishing Group UK London}
}

@article{perdew1996generalized,
  title={Generalized gradient approximation made simple},
  author={Perdew, John P and Burke, Kieron and Ernzerhof, Matthias},
  journal={Physical review letters},
  volume={77},
  number={18},
  pages={3865},
  year={1996},
  publisher={APS}
}

@article{grimme2010consistent,
  title={A consistent and accurate ab initio parametrization of density functional dispersion correction (DFT-D) for the 94 elements H-Pu},
  author={Grimme, Stefan and Antony, Jens and Ehrlich, Stephan and Krieg, Helge},
  journal={The Journal of chemical physics},
  volume={132},
  number={15},
  year={2010},
  publisher={AIP Publishing}
}

@article{campi2023expansion,
  title={Expansion of the materials cloud 2D database},
  author={Campi, Davide and Mounet, Nicolas and Gibertini, Marco and Pizzi, Giovanni and Marzari, Nicola},
  journal={ACS nano},
  volume={17},
  number={12},
  pages={11268--11278},
  year={2023},
  publisher={ACS Publications}
}

@article{san2014electronic,
  title={Electronic structure of spontaneously strained graphene on hexagonal boron nitride},
  author={San-Jose, Pablo and Guti{\'e}rrez-Rubio, A and Sturla, Mauricio and Guinea, Francisco},
  journal={Physical Review B},
  volume={90},
  number={11},
  pages={115152},
  year={2014},
  publisher={APS}
}

@article{nam2017lattice,
  title={Lattice relaxation and energy band modulation in twisted bilayer graphene},
  author={Nam, Nguyen NT and Koshino, Mikito},
  journal={Physical Review B},
  volume={96},
  number={7},
  pages={075311},
  year={2017},
  publisher={APS}
}

@article{pixley2025twisted,
  title={Twisted nodal superconductors},
  author={Pixley, JH and Volkov, Pavel A},
  journal={arXiv preprint arXiv:2503.23683},
  year={2025}
}

@article{can2021high,
  title={High-temperature topological superconductivity in twisted double-layer copper oxides},
  author={Can, Oguzhan and Tummuru, Tarun and Day, Ryan P and Elfimov, Ilya and Damascelli, Andrea and Franz, Marcel},
  journal={Nature Physics},
  volume={17},
  number={4},
  pages={519--524},
  year={2021},
  publisher={Nature Publishing Group UK London}
}

@article{volkov2023current,
  title={Current-and field-induced topology in twisted nodal superconductors},
  author={Volkov, Pavel A and Wilson, Justin H and Lucht, Kevin P and Pixley, JH},
  journal={Physical review letters},
  volume={130},
  number={18},
  pages={186001},
  year={2023},
  publisher={APS}
}

@article{lee2021twisted,
  title={Twisted van der Waals Josephson junction based on a high-T c superconductor},
  author={Lee, Jongyun and Lee, Wonjun and Kim, Gi-Yeop and Choi, Yong-Bin and Park, Jinho and Jang, Seong and Gu, Genda and Choi, Si-Young and Cho, Gil Young and Lee, Gil-Ho and others},
  journal={Nano Letters},
  volume={21},
  number={24},
  pages={10469--10477},
  year={2021},
  publisher={ACS Publications}
}

@article{zhao2023time,
  title={Time-reversal symmetry breaking superconductivity between twisted cuprate superconductors},
  author={Zhao, SY Frank and Cui, Xiaomeng and Volkov, Pavel A and Yoo, Hyobin and Lee, Sangmin and Gardener, Jules A and Akey, Austin J and Engelke, Rebecca and Ronen, Yuval and Zhong, Ruidan and others},
  journal={Science},
  volume={382},
  number={6677},
  pages={1422--1427},
  year={2023},
  publisher={American Association for the Advancement of Science}
}

@misc{yabuki2016supercurrent,
  title={Supercurrent in van der Waals Josephson junction Nat},
  author={Yabuki, N and Moriya, R and Arai, M and Sata, Y and Morikawa, S and Masubuchi, S and Machida, T},
  year={2016},
  publisher={Commun}
}

@article{farrar2021superconducting,
  title={Superconducting quantum interference in twisted van der Waals heterostructures},
  author={Farrar, Liam S and Nevill, Aimee and Lim, Zhen Jieh and Balakrishnan, Geetha and Dale, Sara and Bending, Simon J},
  journal={Nano Letters},
  volume={21},
  number={16},
  pages={6725--6731},
  year={2021},
  publisher={ACS Publications}
}

@article{jian2022superconducting,
  title={Superconducting quantum interference effect in NbSe2/NbSe2 van der Waals junctions},
  author={Jian, Yu and Feng, Qi and Zhong, Jinrui and Peng, Huimin and Duan, Junxi},
  journal={Journal of Physics: Condensed Matter},
  volume={34},
  number={40},
  pages={405702},
  year={2022},
  publisher={IOP Publishing}
}

@article{tani2025twist,
  title={Twist-angle tunable Josephson junctions in three-dimensional superconductors},
  author={Tani, Tenta and Kawakami, Takuto and Koshino, Mikito},
  journal={arXiv preprint arXiv:2508.09551},
  year={2025}
}

@article{xian2019multiflat,
  title={Multiflat bands and strong correlations in twisted bilayer boron nitride: Doping-induced correlated insulator and superconductor},
  author={Xian, Lede and Kennes, Dante M and Tancogne-Dejean, Nicolas and Altarelli, Massimo and Rubio, Angel},
  journal={Nano letters},
  volume={19},
  number={8},
  pages={4934--4940},
  year={2019},
  publisher={ACS Publications}
}

@article{dang2025twisting,
  title={Twisting in h-BN bilayers and their angle-dependent properties},
  author={Dang, Diem Thi-Xuan and Le, Dai-Nam and Woods, Lilia M},
  journal={Physical Review B},
  volume={112},
  number={8},
  pages={085102},
  year={2025},
  publisher={APS}
}

@article{roman2023excitons,
  title={Excitons in twisted AA hexagonal boron nitride bilayers},
  author={Roman-Taboada, Pedro and Obregon-Castillo, Estefania and Botello-Mendez, Andr{\'e}s R and Noguez, Cecilia},
  journal={Physical Review B},
  volume={108},
  number={7},
  pages={075109},
  year={2023},
  publisher={APS}
}

@article{apelian2024delocalization,
  title={Delocalization of Quasiparticle Moir{\'e} States in Twisted Bilayer hBN},
  author={Apelian, Arsineh and Canestraight, Annabelle and Liu, Songyuan and Vlcek, Vojtech},
  journal={Nano Letters},
  volume={24},
  number={38},
  pages={11882--11888},
  year={2024},
  publisher={ACS Publications}
}

@article{yasuda2021stacking,
  title={Stacking-engineered ferroelectricity in bilayer boron nitride},
  author={Yasuda, Kenji and Wang, Xirui and Watanabe, Kenji and Taniguchi, Takashi and Jarillo-Herrero, Pablo},
  journal={Science},
  volume={372},
  number={6549},
  pages={1458--1462},
  year={2021},
  publisher={American Association for the Advancement of Science}
}

@article{yananose2025metamorphic,
  title={Metamorphic quantum dot arrays in twisted trilayer hexagonal boron nitride},
  author={Yananose, Kunihiro and Park, Changwon and Son, Young-Woo},
  journal={arXiv preprint arXiv:2504.14925},
  year={2025}
}

@article{kim2024electrostatic,
  title={Electrostatic moir{\'e} potential from twisted hexagonal boron nitride layers},
  author={Kim, Dong Seob and Dominguez, Roy C and Mayorga-Luna, Rigo and Ye, Dingyi and Embley, Jacob and Tan, Tixuan and Ni, Yue and Liu, Zhida and Ford, Mitchell and Gao, Frank Y and others},
  journal={Nature materials},
  volume={23},
  number={1},
  pages={65--70},
  year={2024},
  publisher={Nature Publishing Group UK London}
}

@article{PhysRevB.108.075109,
  title = {Excitons in twisted {$AA'$} hexagonal boron nitride bilayers},
  author = {Roman-Taboada, Pedro and Obregon-Castillo, Estefania and Botello-Mendez, Andr\'es R. and Noguez, Cecilia},
  journal = {Phys. Rev. B},
  volume = {108},
  issue = {7},
  pages = {075109},
  numpages = {12},
  year = {2023},
  month = {Aug},
  publisher = {American Physical Society},
  doi = {10.1103/PhysRevB.108.075109},
  url = {https://link.aps.org/doi/10.1103/PhysRevB.108.075109}
}

@article{ramos2025flat,
  title={Flat and tunable moir{\'e} phonons in twisted transition-metal dichalcogenides},
  author={Ramos-Alonso, Alejandro and Remez, Benjamin and Bennett, Daniel and Fernandes, Rafael M and Ochoa, H{\'e}ctor},
  journal={Physical review letters},
  volume={134},
  number={2},
  pages={026501},
  year={2025},
  publisher={APS}
}

@article{jung2015origin,
  title={Origin of band gaps in graphene on hexagonal boron nitride},
  author={Jung, Jeil and DaSilva, Ashley M and MacDonald, Allan H and Adam, Shaffique},
  journal={Nature communications},
  volume={6},
  number={1},
  pages={6308},
  year={2015},
  publisher={Nature Publishing Group UK London}
}

@article{sachs2011adhesion,
  title={Adhesion and electronic structure of graphene on hexagonal boron nitride substrates},
  author={Sachs, B and Wehling, TO and Katsnelson, MI and Lichtenstein, AI},
  journal={Physical Review B—Condensed Matter and Materials Physics},
  volume={84},
  number={19},
  pages={195414},
  year={2011},
  publisher={APS}
}

@article{krisna2023moire,
  title={Moir{\'e} phonons in graphene/hexagonal boron nitride moir{\'e} superlattice},
  author={Krisna, Lukas PA and Koshino, Mikito},
  journal={Physical Review B},
  volume={107},
  number={11},
  pages={115301},
  year={2023},
  publisher={APS}
}

@article{nakatsuji2025moire,
  title={Moir{\'e} band engineering in twisted trilayer WSe2},
  author={Nakatsuji, Naoto and Kawakami, Takuto and Tateishi, Hayato and Kato, Koichiro and Koshino, Mikito},
  journal={Communications Materials},
  volume={6},
  number={1},
  pages={274},
  year={2025},
  publisher={Nature Publishing Group UK London}
}

@article{elias2019direct,
  title={Direct band-gap crossover in epitaxial monolayer boron nitride},
  author={Elias, Christine and Valvin, Pierre and Pelini, Thomas and Summerfield, A and Mellor, CJ and Cheng, Tin S and Eaves, Laurence and Foxon, CT and Beton, PhH and Novikov, SV and others},
  journal={Nature communications},
  volume={10},
  number={1},
  pages={2639},
  year={2019},
  publisher={Nature Publishing Group UK London}
}

@article{PhysRevB.105.245408,
  title = {Moir\'e disorder effect in twisted bilayer graphene},
  author = {Nakatsuji, Naoto and Koshino, Mikito},
  journal = {Phys. Rev. B},
  volume = {105},
  issue = {24},
  pages = {245408},
  numpages = {11},
  year = {2022},
  month = {Jun},
  publisher = {American Physical Society},
  doi = {10.1103/PhysRevB.105.245408},
  url = {https://link.aps.org/doi/10.1103/PhysRevB.105.245408}
}

@article{ghorashi2023topological,
  title={Topological and stacked flat bands in bilayer graphene with a superlattice potential},
  author={Ghorashi, Sayed Ali Akbar and Dunbrack, Aaron and Abouelkomsan, Ahmed and Sun, Jiacheng and Du, Xu and Cano, Jennifer},
  journal={Physical Review Letters},
  volume={130},
  number={19},
  pages={196201},
  year={2023},
  publisher={APS}
}

@article{ghorashi2023multilayer,
  title={Multilayer graphene with a superlattice potential},
  author={Ghorashi, Sayed Ali Akbar and Cano, Jennifer},
  journal={Physical Review B},
  volume={107},
  number={19},
  pages={195423},
  year={2023},
  publisher={APS}
}

@article{zeng2024gate,
  title={Gate-tunable topological phases in superlattice modulated bilayer graphene},
  author={Zeng, Yongxin and Wolf, Tobias MR and Huang, Chunli and Wei, Nemin and Ghorashi, Sayed Ali Akbar and MacDonald, Allan H and Cano, Jennifer},
  journal={Physical Review B},
  volume={109},
  number={19},
  pages={195406},
  year={2024},
  publisher={APS}
}

@article{guerci2022designer,
  title={Designer meron lattice on the surface of a topological insulator},
  author={Guerci, Daniele and Wang, Jie and Pixley, JH and Cano, Jennifer},
  journal={Physical Review B},
  volume={106},
  number={24},
  pages={245417},
  year={2022},
  publisher={APS}
}

@article{seleznev2024inducing,
  title={Inducing topological flat bands in bilayer graphene with electric and magnetic superlattices},
  author={Seleznev, Daniel and Cano, Jennifer and Vanderbilt, David},
  journal={Physical Review B},
  volume={110},
  number={20},
  pages={205115},
  year={2024},
  publisher={APS}
}

@article{ault2025optimizing,
  title={Optimizing superlattice bilayer graphene for a fractional Chern insulator},
  author={Ault-McCoy, Dathan and Lhachemi, M Nabil Y and Dunbrack, Aaron and Ghorashi, Sayed Ali Akbar and Cano, Jennifer},
  journal={arXiv preprint arXiv:2505.05551},
  year={2025}
}

@article{morales2023pressure,
  title={Pressure-enhanced fractional Chern insulators along a magic line in moir{\'e} transition metal dichalcogenides},
  author={Morales-Dur{\'a}n, Nicol{\'a}s and Wang, Jie and Schleder, Gabriel R and Angeli, Mattia and Zhu, Ziyan and Kaxiras, Efthimios and Repellin, C{\'e}cile and Cano, Jennifer},
  journal={Physical Review Research},
  volume={5},
  number={3},
  pages={L032022},
  year={2023},
  publisher={APS}
}

@misc{kaplan2025,
  title={in preparation},
  author={Kaplan, Daniel and Tyler, Alex and Pixley, Jedediah}
}

@article{bao2024deep,
  title={Deep-Learning Database of Density Functional Theory Hamiltonians for Twisted Materials},
  author={Bao, Ting and Xu, Runzhang and Li, He and Gong, Xiaoxun and Tang, Zechen and Fu, Jingheng and Duan, Wenhui and Xu, Yong},
  journal={arXiv preprint arXiv:2404.06449},
  year={2024}
}

@article{ashton2017topology,
  title={Topology-scaling identification of layered solids and stable exfoliated 2D materials},
  author={Ashton, Michael and Paul, Joshua and Sinnott, Susan B and Hennig, Richard G},
  journal={Physical review letters},
  volume={118},
  number={10},
  pages={106101},
  year={2017},
  publisher={APS}
}

@article{zhou20192dmatpedia,
  title={2DMatPedia, an open computational database of two-dimensional materials from top-down and bottom-up approaches},
  author={Zhou, Jun and Shen, Lei and Costa, Miguel Dias and Persson, Kristin A and Ong, Shyue Ping and Huck, Patrick and Lu, Yunhao and Ma, Xiaoyang and Chen, Yiming and Tang, Hanmei and others},
  journal={Scientific data},
  volume={6},
  number={1},
  pages={86},
  year={2019},
  publisher={Nature Publishing Group UK London}
}

@article{giannozzi2009quantum,
  title={QUANTUM ESPRESSO: a modular and open-source software project for quantumsimulations of materials},
  author={Giannozzi, Paolo and Baroni, Stefano and Bonini, Nicola and Calandra, Matteo and Car, Roberto and Cavazzoni, Carlo and Ceresoli, Davide and Chiarotti, Guido L and Cococcioni, Matteo and Dabo, Ismaila and others},
  journal={Journal of physics: Condensed matter},
  volume={21},
  number={39},
  pages={395502},
  year={2009},
  publisher={IOP Publishing}
}

\newpage 

\renewcommand{\thefigure}{S\arabic{figure}} 
\setcounter{figure}{0} 
\renewcommand{\thetable}{S\arabic{table}} 
\setcounter{table}{0} 

\section*{Methods}

We now describe the technical details of the calculations underlying the workflow presented in Fig.~\ref{fig_workflow}. 

\paragraph*{Monolayers:} We begin by enumerating the twistable semiconducting monolayers with hexagonal lattice and electrostatically dopable pockets at $K/K'$, as identified in Ref.~\cite[App.~IV-2.c and~IV-2.d]{jiang20242d} by scanning the C2DB~\cite{haastrup2018computational,campi2023expansion} and MC2D~\cite{mounet2018two}. 
Amongst the 887 reported $K$-valley semiconductors, we kept the 141 that have $K$-valley pockets, we found a few were misclassified and we added BN leading to a total of 133 quoted in the text.
For each material, we retrieve the structure file and perform a relaxed self-consistent field (SCF) calculation on a $16\times 16$ k-point mesh as a reference for subsequent commensurate bilayer calculations. All the density functional theory (DFT) calculations used to obtain the results in the main text employ the PBE exchange-correlation functional~\cite{perdew1996generalized} together with van der Waals corrections using the D3 method of Grimme \textit{et al.}~\cite{grimme2010consistent}. Monolayers were fully relaxed prior to the bilayer construction, while atoms were relaxed only along the $z$-direction in bilayer calculations. 

\paragraph*{Moir\'e bilayers:} To construct the twisted homobilayers studied in the main text, we duplicate the monolayer, apply a mirror-symmetry operation with respect to the axis perpendicular to the layer plane ($M_z$), rotate it by an angle $\theta$, and stack the mirrored-and-rotated copy atop the original monolayer
For many materials, such as traditional TMDs (\textit{e.g.} MoTe$_2$), the application of $M_z$ is trivial; however, this procedure allows us to systematically treat materials lacking $M_z$ symmetry, such as Janus TMDs (\textit{e.g.} MoSeTe). 
Indeed, irrespective of the monolayer, the resulting bilayer structure possesses all symmetries assumed in the continuum model originally derived for MoTe$_2$~\cite{wu2019topological}
\begin{subequations} \label{eq_continuum} \begin{align}
&H (d) = \begin{bmatrix} \frac{(k-\kappa_t)^2}{2m^*} + \Delta_+(d) & T(d) \\  T^*(d) &  \frac{(k-\kappa_b)^2}{2m^*} + \Delta_{-}(d) \end{bmatrix} ,  \\ 
&\Delta_\pm(r) = 2 v \sum_{j=1,3,5} \cos(G_j \cdot d \pm \psi) , \\ 
&T(d) = w (1+e^{-i G_2\cdot d}+e^{-i G_3\cdot d} ) , 
\end{align} \end{subequations}
with $\kappa_{t/b}$ the two inequivalent corners of the moiré Brillouin zone on which the $K$-point of the top/bottom layer downfold, $d(r)$ representing the interlayer displacement vector at position $r$ in the moiré unit cell, and where we follow and refer to the notations of Refs.~\cite{wu2019topological,devakul2021magic}. 
This makes Eq.~\ref{eq_continuum} applicable to all studied bilayers, for some values of the parameters: $m^*$ (effective mass, negative for valence band pockets), $v$ (intralayer potential amplitude), $w$ (interlayer tunneling amplitude) and $\psi$ (intralayer potential phase). 
We truncate the Hamiltonian to the first shell of monolayer reciprocal lattice vectors, the $G_j$ in Eq.~\ref{eq_continuum} labeled clockwise as $j$ runs from 1 to 6~\cite{devakul2021magic}. Higher harmonics contribute only at subleading order in our perturbative treatment, and we omit them here for clarity.

\paragraph*{Parameter estimation:} We obtain the effective mass from the relaxed DFT calculation of the monolayer. To fix the other parameters, we perform relaxed DFT calculations for commensurate bilayers ($\kappa_t = \kappa_b$) for three particular displacement vectors $d_{\rm AA}$, $d_{\rm AB}$ and $d_{\rm BA}$ respectively satisfying $d_{\rm AA} \cdot G_{j=1,3,5} = 0$, $d_{\rm AB} \cdot G_{j=1,3,5} = 2\pi/3$ and $d_{\rm BA} \cdot G_{j=1,3,5} = 4\pi/3$. Looking at the energy band splitting $\delta \varepsilon_{{\rm AA}, {\rm AB}, {\rm BA}}$ at $k=\kappa_t=\kappa_b$, we infer the moiré parameters by comparing with the prediction of Eq.~\ref{eq_continuum}:
\begin{subequations}
\begin{align}
\delta \varepsilon_{{\rm AA}} & = 3 w,\\
\delta \varepsilon_{{\rm AB}} & = 2v\cos(\psi + 2\pi/3),\\
\delta \varepsilon_{{\rm BA}} & = 2v\cos(\psi - 2\pi/3),
\end{align}
\end{subequations}
The extracted parameters are listed in Tabs.~\ref{tab_listparms}~-~\ref{tab_listparms_soc}.

At this stage, we discard materials for which the moiré coupling exceeds the monolayer interband gap, as this would require including multiple bands beyond the quadratic minimum/maximum in our low-energy description (monolayer kinetic terms in Eq.~\ref{eq_continuum}). The materials discarded feature ``nan'' values for $(v,w,\psi)$ in Tab.~\ref{tab_listparms}. 
This exclusion does not imply these materials are unsuitable for moiré physics. For example, BSb and GaSe were removed from the database of Ref.~\cite{jiang20242d} because they exhibit strong interlayer coupling that significantly alters their band structures. Nevertheless, this strong coupling suggests these systems could be promising candidates for engineering highly modulated moiré bands. In fact, they appear as leading candidates in Fig.~\ref{fig_polar_field} for producing strong alternating electric fields at polar domains. Developing accurate models for such materials will be the focus of future work.

\paragraph*{Perturbative moiré indicators:} The perturbative moiré indicators were worked out for the model Eq.~\ref{eq_continuum} in Ref.~\cite[App.~C]{crepel2025efficient}, where formulas for the Chern number of the twisted bilayer in the large twist angle limit and the perturbative gap opened by the long-wavelength moiré fields were derived as a function of $v$, $w$ and $\psi$. We use these results to obtain Fig.~\ref{fig_fullclassification}a-b.

\paragraph*{Magic-angle estimation:} To quickly identify the magic angle, we solve the continuum model in Eq.~\ref{eq_continuum} truncated to the first shell~\cite{bistritzer2011moire,devakul2021magic} at the center $\gamma$ and corner $\kappa_t$ of the moiré Brillouin zone. 
A reliable proxy for the emergence of a flat band is when the energies of the first moiré band at these two points become equal, forcing the band to minimize its bandwidth to interpolate between identical values. 
These energies read 
\begin{align}
E_\gamma (\theta) & = e_0 + {\rm mm}_n [ 2w \cos(\pi n/3) + 2v \cos(2\pi n/3 - \psi) ] , \notag \\
E_\kappa (\theta) & = {\rm mm} \left[e_+ \pm   \sqrt{e_+^2 + 3 w^2} , e_0 + 2 v \cos  \left( \psi \pm \frac{2\pi}{3} \right) \right] , 
\end{align} 
with $e_0 = |\kappa_t|^2 / (2m^*)$ and $e_+ = (e_0 + 2v\cos\psi)/2$; and where ${\rm mm} = {\rm min/max}$ for conduction/valence band pockets. Solving $E_\gamma (\theta_m) = E_\kappa (\theta_m)$ yields the magic angle proxy $\theta_m$, plotted as stars in Fig.~\ref{fig_fullclassification}c-d. 

\paragraph*{Lattice relaxation threshold:} 
The strength of in-plane lattice relaxation in twisted homobilayers can be characterized by the dimensionless parameter $\eta = U_{\rm vdw} / \mu\theta^{2}$, with $\mu$ the Lamé elastic coefficient and $U_{\rm vdw}$ the interlayer van der Waals energy scale~\cite{nam2017lattice,ezzi2024analytical}. The elastic coefficients can be deduced from the lattice stiffness coefficients provided for all materials in the C2DB database~\cite{haastrup2018computational,campi2023expansion}. 
The interlayer van der Waals energy is obtained by fitting the total energies of our commensurate bilayers obtained by DFT with the formula~\cite{nam2017lattice,carr2018relaxation,ezzi2024analytical}
\begin{equation} \label{appeq_energyvdw}
U_{\rm b} (d) = U_{0} + 2U_{\rm vdw}\sum_{j=1,3,5} \cos\left(G_j \cdot d\right) ,
\end{equation}
with $U_0$ a global constant. The variables $U_0$ and $U_{\rm vdw}$ are fully fixed by the total DFT energies obtained for $d_{\rm AA}$ and $d_{\rm AB}$. This coefficient is listed for all materials in Tabs.~\ref{tab_listparms}~-~\ref{tab_listparms_soc}. It directly enables the inclusion of lattice relaxation effect in the continuum model of Eq.~\ref{eq_continuum}, either through the analytic approximation of Refs.~\cite{ezzi2024analytical,de2025theory}, or by minimization of the elasticity energy functional that we now briefly introduce.

\begin{figure}
\centering
\includegraphics[width=\columnwidth]{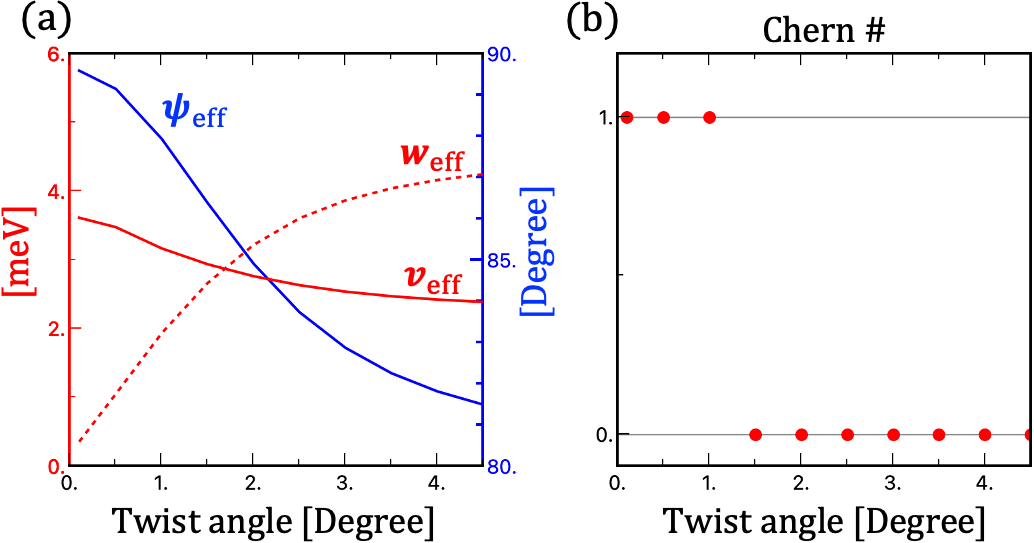}
\caption{\textit{
(a) 1st harmonic of the continuum model renormalized by in-plane relaxation, obtained by minimization of the bilayer lattice energy as a function of atomic displacements for ZrBr$_2$. The red solid line, red dashed line, and blue line represent 
$v_{\rm eff}$, $w_{\rm eff}$, and $\psi_{\rm eff}$, respectively. 
The left vertical axis shows the energy scale for $v_{\rm eff}$ and $w_{\rm eff}$ in meV, while the right vertical axis shows the angle (in degrees) for $\psi_{\rm eff}$. 
(b) Chern number obtained from perturbation theory; a transition to a topological band is observed as the twist angle is reduced below $1^\circ$.
}}
\label{fig_renormalized_parameters}
\end{figure}

\paragraph*{Renormalization of harmonics by in-plane relaxation:} We give a brief overview of the bilayer elastic theory necessary to understand the relaxation-induced topological transition, refering to Refs.~\cite{nam2017lattice,nakatsuji2025moire} for more details. 
For in-plane atomic displacements $u^{(\ell=1/2)}$ in the top/bottom layer, the monolayer elastic energy 
\begin{align}
U_{\rm ela}=\sum_{\ell=1}^{2} \frac{1}{2}&\left[\left(\mu+\lambda\right)\left(u_{xx}^{(\ell)}+u_{yy}^{(\ell)}\right)^{2} \right.\notag \\
&\left. +\mu\left\{\left(u_{xx}^{(\ell)}-u_{yy}^{(\ell)}\right)^{2}+4\left(u_{xy}^{(\ell)}\right)^{2}\right\}\right],
\end{align}
and van der Waals energy of Eq.~\ref{appeq_energyvdw} combine into the total energy functional of the system $E_{\rm tot} = \int {\rm d}r^2 (U_{\rm ela} + U_{\rm b})$, where we have also introduced strain tensor $u_{ij}^{(\ell)}=(\partial_{i}u_{j}^{(\ell)}+\partial_{j}u_{i}^{(\ell)})/2$.
Minimization of $E_{\rm tot}$ gives the optimized moiré structure $u^{(\ell)}(r)$ after in-plane lattice relaxation. 
We inject this solution into the continuum model through the replacement $d(r) \to \tilde{d}(r) = \theta \hat{z} \times r + u^{(2)}-u^{(1)}$, which leads to 
renormalized moiré fields
\begin{align} \label{eq_relaxed_continuum}
    &\Delta_\pm(r) = 2 v \sum_{j=1,3,5} \cos(G_j \cdot \tilde{d} \pm \psi) , \nonumber \\ 
    &T(r) = w \sum_{j=1} e^{i Q_j\cdot\tilde{d}}, 
\end{align}
with $Q_j = C_3^j K$ the three $K$-points of the monolayers before rotation, and where we have performed a gauge transformation prior to the substitution of $\tilde{d}$ in the continuum model to recover the form given in Ref.~\cite{nakatsuji2025moire}. 
Expanding Eq.~\ref{eq_relaxed_continuum} in moiré harmonics, the first harmonic terms give renormalized parameters $w_{\rm eff}$, $v_{\rm eff}$, and $\psi_{\rm eff}$.
Fig.~\ref{fig_renormalized_parameters}a gives the twist angle dependence of the renomalized parameters for ZrBr$_2$. Lattice relaxation reduces the interlayer coupling $w_{\rm eff}$ and enhances the intralayer potential $v_{\rm eff}$ compared to their respective bare values $w$ and $v$. This relaxation-induced modulation drives the emergence of a nonzero Chern number around a twist angle of $1.5^\circ$, as shown in Fig.~\ref{fig_renormalized_parameters}b. The underlying
mechanism is that relaxation suppresses the AA-stacking regions, where interlayer
hopping is strongest. Consequently, this type of parameter renormalization is
expected to be a generic phenomenon in materials with positive $U_{\mathrm{vdW}}$.

\paragraph*{Robustness to computational choice:}

\begin{figure*}
\centering
\includegraphics[width=\textwidth]{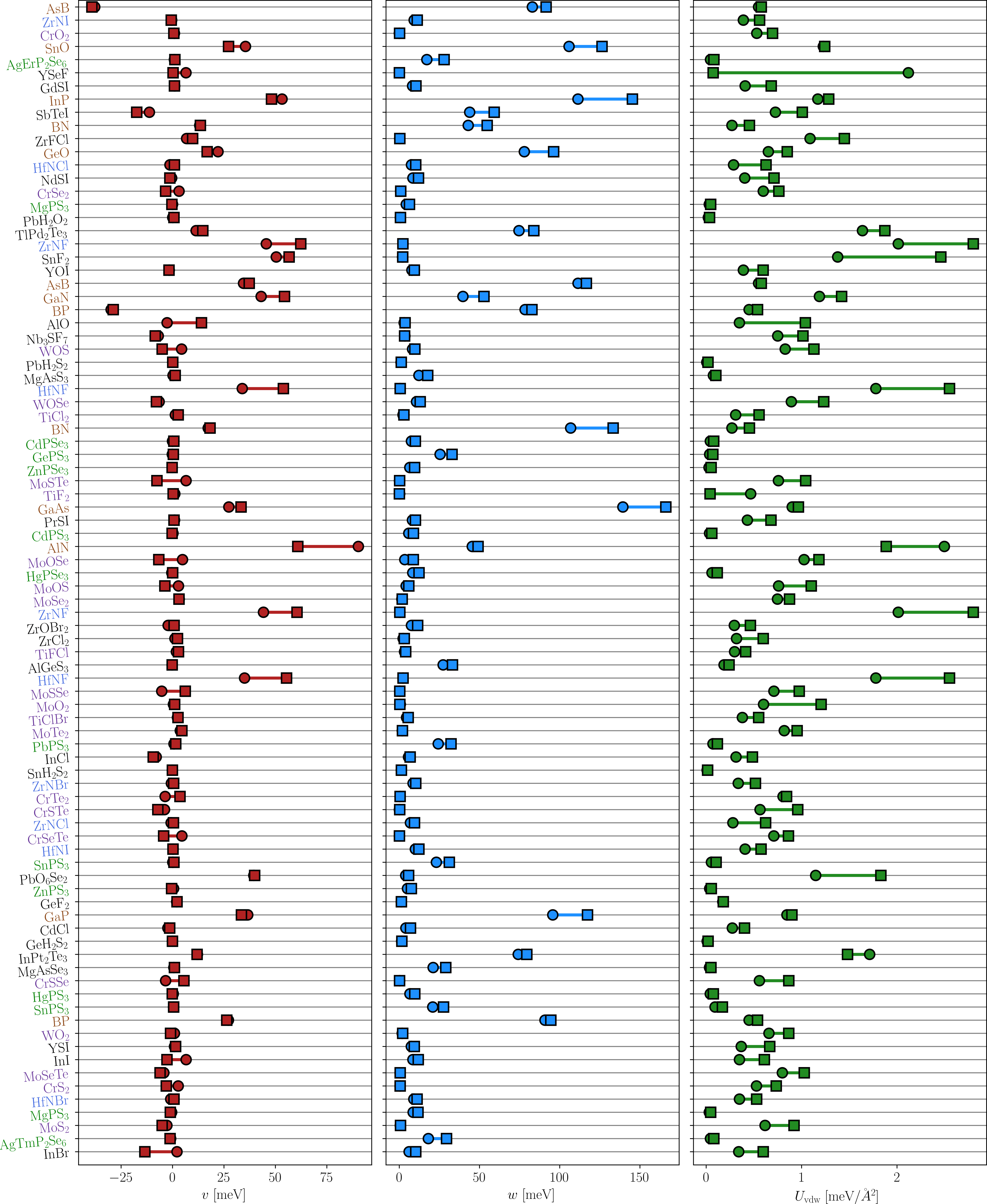}
\caption{\textit{
Comparison of the parameters $v$, $w$ and $U_{\rm vdw}$ extracted from our DFT calculations for all materials without spin-orbit coupling using either D3 (circles) and rvv10 (squares) van der Waals correction functionals. The colors of the material names follow those in Fig.~\ref{fig_fullclassification}. 
}}
\label{fig_functionals}
\end{figure*}

We verified the robustness of our results against variations in the DFT functional by comparing the DFT parameters extracted with both D3 and rVV10 van der Waals corrections. As shown in Fig.~\ref{fig_functionals}, the values of $v$ and $w$ display good quantitative agreement across van der Waals correction functionals. This underscores the consistency of our perturbative approach and the validity of the resulting predictions for the moiré gap and Chern number of the homobilayers, reported in Fig.~\ref{fig_fullclassification} of the main text. 
The choice of van der Waals correction has a stronger effect on the relaxation properties of the bilayer, as can be seen in Fig.~\ref{fig_functionals}, where we also show the van der Waals energy $U_{\rm vdw}$ (see text and definition below). Unlike the continuum model parameters, this coefficient exhibits larger variations across functionals -- an expected feature, as it directly depends on the accuracy of the van der Waals force description in DFT. Consequently, the relaxation angles extracted from it (right panels of Fig.~\ref{fig_fullclassification}a-b) is subject to greater uncertainty.

\begin{figure*}
\centering
\includegraphics[width=\textwidth]{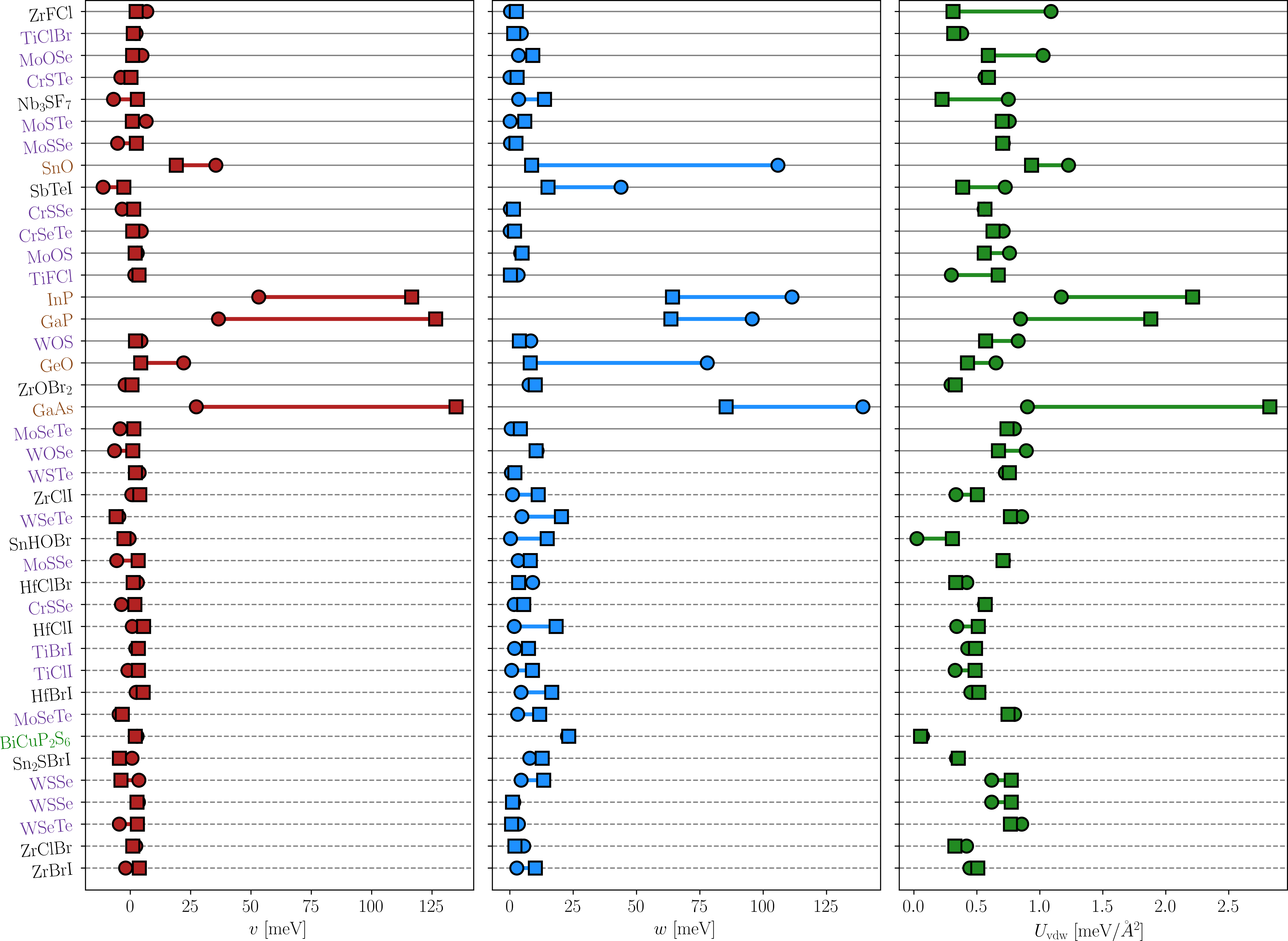}
\caption{\textit{
Comparison of the parameters $v$, $w$ and $U_{\rm vdw}$ extracted from our DFT calculations for monolayers lacking $M_z$-symmetry in the two inequivalent $E/M_z$ (circles) and $M_z/E$ (squares) stacking configurations. The colors of the material names and linestyle follow those in Fig.~\ref{fig_fullclassification}. 
}}
\label{fig_stacking}
\end{figure*}

\paragraph*{$E/M_z$ versus $M_z/E$ configurations:} For monolayers lacking $M_z$-symmetry, an additional arbitrary choice in our calculations is whether the $M_z$-flip is applied to the top or bottom layer (see above). We refer to these configurations as $E/M_z$ and $M_z/E$, respectively. Concretely, for the Janus TMD MoSSe with vertical atomic alignment SMoSe, this degree of freedom distinguishes the two inequivalent bilayers SeMoS–SMoSe ($E/M_z$) and SMoSe–SeMoS ($M_z/E$). 
To assess the sensitivity of our results to this choice, we compare the DFT-extracted parameters $v$, $w$ and $U_{\rm vdw}$ for both configurations across all monolayers without $M_z$-symmetry. As shown in Fig.~\ref{fig_stacking}, the stacking choice generally leads to only minor renormalization of the continuum-model coefficients. The only exceptions are one-atom-thick binary materials (SnO, GeO, InP, GaP, GaAs), where strong honeycomb buckling and orbital localization on a single atomic species significantly enhance coupling when these atoms are closer together. 
For example, in GaP (highlighted in Fig.~\ref{fig_bands}c),  the valence band is dominated by P-$p$ orbitals~\cite{haastrup2018computational}, making the interlayer tunneling $w$ stronger in the GaP-PGa ($E/M_z$) configuration compared to the PGa-GaP ($M_z/E$). 
For the five monolayers where the stacking choice produces large parameter variations, we selected the stacking configuration yielding the largest interlayer tunneling $w$ for Fig.~\ref{fig_fullclassification} -- corresponding to the $E/M_z$ class, identified in Tabs.~\ref{tab_liststack}~-~\ref{tab_liststacksoc}.

\section*{Table of model parameters}

Finally, we gather in Tabs.~\ref{tab_listparms}~-~\ref{tab_listparms_soc} all the parameters and coefficients necessary to describe the bilayer continuum models, their relaxation properties within the elastic theory described above, and other properties needed to reproduce the results presented in the main text. 
Tab.~\ref{tab_listparms} and~\ref{tab_listparms_soc} respectively contain the materials without and with spin-orbit coupling. 

In each table, the first four columns specify the monolayer considered: the first gives the material name, the second refers to whether the coefficients are given for the conduction or valence band pockets (this s also given by the sign of the effective mass), the third gives the corresponding material ID in the database Ref.~\cite{petralanda2024two} that uniquely characterizes the material when the same chemical composition can crystallize in different lattices, the fourth is the global (indirect) gap $E_g$ in eV, giving a proxy for how easily dopable these semiconducting materials are. 
The columns 5-9 provides the coefficients defined in Eq.~\ref{eq_continuum} obtained following the prescription detailed in the Methods section, leaving ``nan'' when Eq.~\ref{eq_continuum} cannot be used (see discussion above). They respectively show the monolayer lattice constant (fifth), the effective mass (sixth), the moiré interlayer hopping amplitude (seventh), the moiré intralayer potential amplitude (eighth) and shift (ninth). 
To uniquely define the parameters, we choose a gauge in which $w>0$ and, without loss of generality, take $0^\circ\leq\psi<90^\circ$ (other values of $\psi$ are equivalent due to mirror symmetry and up to the change $v \leftrightarrow -v$).
The last three columns summarize the parameters needed to characterize relaxation in the bilayer. Columns 10-11 list the $(\lambda,\mu)$-Lamé coefficients extracted from the C2DB database (for BN, we use the values of Ref.~\cite{jung2015origin,sachs2011adhesion,krisna2023moire}), leaving ``nan'' when the data is not available; and the second to last column is the interlayer van der Waals energy. Finally, the last column shows the electric field at the AB stacking.

\onecolumngrid 
\centering 

\begin{longtable}{c|c|c|c||c|c|c|c|c|c|c|c|c}
\caption{\textit{
Parameters for materials without spin-orbit coupling: $E/M_z$.
}}\label{tab_listparms}\\
\toprule
Name & band & ID & $E_g$ [eV]& $a_0$ [$\AA$] & $\frac{m^*}{m_e}$ & $w$ [meV] & $v$ [meV] & $\psi$ [$^\circ$] & $\lambda$ $\left[\frac{\rm eV}{\AA^2}\right]$ & $\mu$ $\left[\frac{\rm eV}{\AA^2}\right]$ & $U_{\rm vdw}$ $\left[\frac{\rm meV}{\AA^2}\right]$ & $\delta_{\rm AB}$ $\left[\frac{\rm meV}{\AA}\right]$ \\[6pt]\hline
\midrule
\endfirsthead
\midrule
\endfoot
\bottomrule
\endlastfoot
ZrBrN & conduction & 6.3.2323 & 1.62 & 3.642 & 0.578 & 8.58333 & -0.56626 & 57.263 & 10.611 & 2.854 & 0.336 & 0.144 \\ \hline
ZrClN & conduction & 6.3.2408 & 1.91 & 3.614 & 0.576 & 6.95000 & -0.69534 & 4.763 & 11.555 & 2.731 & 0.280 & -0.070 \\ \hline
MoS2 & conduction & 3.1.39 & 1.6 & 3.164 & 0.486 & 0.63333 & -2.73780 & 89.315 & 3.073 & 3.072 & 0.617 & -2.004 \\ \hline
MoSe2 & conduction & 3.1.41 & 1.34 & 3.282 & 0.601 & 1.60000 & 3.25401 & 84.309 & 2.197 & 2.623 & 0.747 & -2.552 \\ \hline
MoTe2 & conduction & 3.1.43 & 0.96 & 3.523 & 0.610 & 1.86667 & 3.77414 & 73.723 & 1.680 & 1.992 & 0.818 & -3.259 \\ \hline
GaN & valence & 6.1.78 & 1.82 & 3.204 & -1.224 & 39.66667 & 43.06365 & 73.567 & 171.895 & 2.199 & 1.187 & -26.072 \\ \hline
MoSSe & conduction & 3.1.10 & 1.48 & 3.223 & 0.560 & 0.26667 & -5.25230 & 86.831 & 2.548 & 2.850 & 0.710 & -3.617 \\ \hline
PbO6Se2 & conduction & 6.2.1287 & 0.49 & 5.664 & 0.355 & 4.00000 & 39.49689 & 83.719 & nan & nan & 1.148 & 22.754 \\ \hline
SnPS3 & conduction & 3.2.152 & 1.59 & 6.624 & 0.645 & 20.85000 & 0.51606 & 55.127 & nan & nan & 0.093 & 0.336 \\ \hline
AgErP2Se6 & conduction & 3.2.153 & 1.5 & 6.556 & 1.096 & 17.16667 & 0.90350 & 80.112 & nan & nan & 0.045 & 0.572 \\ \hline
AgTmP2Se6 & conduction & 3.2.158 & 1.5 & 6.545 & 1.114 & 18.11667 & -0.92567 & 76.397 & nan & nan & 0.045 & 0.545 \\ \hline
YFSe & conduction & 6.2.1225 & 2.57 & 4.790 & 0.601 & 0.06667 & 6.53321 & 13.848 & nan & nan & 2.119 & 5.677 \\ \hline
PrIS & conduction & 6.2.1268 & 2.47 & 4.656 & 1.019 & 8.20000 & 0.97100 & 66.444 & nan & nan & 0.432 & 0.570 \\ \hline
NdIS & conduction & 6.2.1266 & 2.47 & 4.617 & 1.084 & 8.48333 & -0.43162 & 47.367 & nan & nan & 0.404 & 0.230 \\ \hline
GdIS & conduction & 6.2.1238 & 2.57 & 4.491 & 1.580 & 8.28333 & 1.03394 & 33.305 & nan & nan & 0.408 & -0.293 \\ \hline
BP & conduction & 6.3.2606 & 0.91 & 3.214 & 0.210 & 78.41667 & -29.84890 & 76.478 & 3.927 & 3.318 & 0.450 & -7.236 \\ \hline
BP & valence & 6.3.2606 & 0.91 & 3.214 & -0.197 & 90.81667 & 27.15441 & 64.169 & 3.927 & 3.318 & 0.450 & -7.236 \\ \hline
TiF2 & valence & 3.3.500 & 1.28 & 2.836 & -1.033 & 0.25000 & 1.10254 & 87.863 & 1.761 & 2.988 & 0.466 & 0.663 \\ \hline
ZrCl2 & valence & 3.3.495 & 0.99 & 3.389 & -0.449 & 2.40000 & 1.16322 & 79.868 & 1.158 & 2.126 & 0.318 & -0.925 \\ \hline
InBr & conduction & 6.3.2321 & 1.34 & 4.080 & 0.382 & 6.11667 & 2.16277 & 2.678 & 2.307 & 0.169 & 0.341 & -0.228 \\ \hline
InI & conduction & 6.3.2500 & 1.01 & 4.324 & 0.391 & 8.48333 & 6.59949 & 15.779 & 0.827 & 0.233 & 0.349 & 1.079 \\ \hline
AlO & valence & 3.3.475 & 1.32 & 2.960 & -0.397 & 3.03333 & -2.64799 & 78.858 & 10.388 & 3.425 & 0.348 & 1.675 \\ \hline
TiCl2 & valence & 3.3.494 & 0.9 & 3.260 & -0.636 & 2.33333 & 1.36327 & 81.180 & 1.425 & 2.103 & 0.309 & -0.839 \\ \hline
InCl & conduction & 6.3.2406 & 1.55 & 3.875 & 0.363 & 5.75000 & -7.93317 & 3.025 & -0.934 & 0.121 & 0.313 & -1.044 \\ \hline
SnH2S2 & valence & 6.3.2485 & 1.0 & 4.579 & -0.871 & 1.26667 & -0.07709 & 87.006 & nan & nan & 0.010 & 0.066 \\ \hline
GeH2S2 & valence & 6.3.2470 & 1.05 & 4.257 & -0.791 & 1.46667 & -0.05774 & 89.460 & nan & nan & 0.011 & 0.149 \\ \hline
AsB & conduction & 6.3.2603 & 0.75 & 3.391 & 0.179 & 82.93333 & -37.81631 & 72.279 & 3.531 & 2.853 & 0.549 & -5.613 \\ \hline
AsB & valence & 6.3.2603 & 0.75 & 3.391 & -0.173 & 111.41667 & 34.50325 & 61.178 & 3.531 & 2.853 & 0.549 & -5.613 \\ \hline
CdCl & valence & 3.3.408 & 1.68 & 3.836 & -0.985 & 4.10000 & -2.27119 & 21.629 & 0.434 & 0.540 & 0.274 & -0.484 \\ \hline
SnO & valence & 6.3.2139 & 1.68 & 3.338 & -1.043 & 105.81667 & 35.45471 & 32.755 & 1.698 & 0.804 & 1.228 & 6.754 \\ \hline
SnPS3 & conduction & 3.3.356 & 1.06 & 6.694 & 0.401 & 23.11667 & 0.10914 & 0.000 & nan & nan & 0.055 & -0.009 \\ \hline
MoO2 & conduction & 3.3.512 & 0.92 & 2.813 & 0.445 & 0.63333 & 0.42280 & 45.800 & 12.133 & 4.665 & 0.601 & 0.099 \\ \hline
TiClF & valence & 3.3.271 & 1.13 & 3.075 & -0.985 & 3.26667 & 1.86171 & 69.736 & nan & nan & 0.298 & -0.982 \\ \hline
ZrClF & valence & 3.3.272 & 1.25 & 3.243 & -0.606 & 0.26667 & 6.83300 & 87.632 & nan & nan & 1.089 & 4.660 \\ \hline
PbH2O2 & valence & 6.3.2479 & 1.68 & 3.996 & -1.541 & 0.66667 & 0.09689 & 57.586 & 5.177 & 0.506 & 0.023 & -0.087 \\ \hline
WO2 & conduction & 3.3.520 & 1.34 & 2.819 & 0.383 & 1.81667 & 0.91928 & 54.216 & 10.904 & 5.478 & 0.658 & 0.292 \\ \hline
GePS3 & conduction & 3.3.331 & 1.42 & 6.361 & 0.428 & 25.43333 & -0.11836 & 0.000 & -0.508 & 0.195 & 0.036 & 0.167 \\ \hline
ZrBr2O & conduction & 6.3.1978 & 1.07 & 3.662 & 0.618 & 7.58333 & -2.14083 & 17.257 & -13.611 & 1.397 & 0.295 & 0.054 \\ \hline
PbPS3 & conduction & 3.3.347 & 1.48 & 6.792 & 0.418 & 24.26667 & 0.70284 & 0.000 & -0.547 & 0.198 & 0.071 & 0.014 \\ \hline
PbH2S2 & valence & 6.3.2484 & 1.47 & 4.671 & -1.109 & 1.10000 & 0.07157 & 70.253 & -0.395 & 0.126 & 0.006 & -0.119 \\ \hline
YIO & conduction & 6.3.2503 & 1.87 & 3.920 & 0.862 & 7.86667 & -1.60790 & 44.202 & 3.316 & 2.354 & 0.390 & -0.279 \\ \hline
MoOS & conduction & 3.3.283 & 1.09 & 2.985 & 0.689 & 4.26667 & 2.89205 & 88.947 & 5.459 & 3.823 & 0.759 & 1.567 \\ \hline
WOSe & conduction & 3.3.290 & 1.26 & 3.043 & 0.903 & 10.73333 & -6.51056 & 86.736 & 3.933 & 4.242 & 0.893 & 3.277 \\ \hline
TiBrCl & valence & 3.3.266 & 0.83 & 3.353 & -0.654 & 4.53333 & 2.39009 & 83.946 & 1.343 & 1.953 & 0.379 & -1.346 \\ \hline
HfBrN & conduction & 6.3.2318 & 1.97 & 3.585 & 0.654 & 9.05000 & -0.89447 & 30.730 & 17.298 & 2.945 & 0.349 & -0.125 \\ \hline
GeO & valence & 6.3.2066 & 2.09 & 2.992 & -1.226 & 77.93333 & 22.10404 & 43.406 & 1.926 & 1.104 & 0.652 & 5.757 \\ \hline
ISbTe & conduction & 6.3.2116 & 1.03 & 4.175 & 0.541 & 43.88333 & -11.24932 & 88.221 & -8.540 & 0.394 & 0.726 & -3.039 \\ \hline
MoOSe & conduction & 3.3.284 & 0.78 & 3.046 & 0.981 & 3.43333 & 4.84814 & 79.255 & 4.469 & 3.557 & 1.026 & 2.630 \\ \hline
SnF2 & valence & 6.3.2432 & 2.36 & 3.863 & -1.496 & 2.05000 & 50.41561 & 86.864 & nan & nan & 1.378 & 23.730 \\ \hline
Nb3F7S & conduction & 6.3.2051 & 0.96 & 6.155 & 3.791 & 3.48333 & -6.94163 & 82.988 & 2.540 & 1.455 & 0.750 & -1.301 \\ \hline
HfIN & conduction & 6.3.2492 & 0.62 & 3.673 & 0.718 & 10.00000 & 0.27577 & 19.359 & 10.513 & 3.288 & 0.408 & -0.028 \\ \hline
BSb & conduction & 6.3.2607 & 0.3 & 3.732 & 0.105 & nan & nan & nan & 3.518 & 2.074 & 2.374 & -26.357 \\ \hline
BSb & valence & 6.3.2607 & 0.3 & 3.732 & -0.104 & nan & nan & nan & 3.518 & 2.074 & 2.374 & -26.357 \\ \hline
HfClN & conduction & 6.3.2403 & 2.33 & 3.548 & 0.639 & 7.58333 & -1.25932 & 1.970 & 26.000 & 2.707 & 0.287 & 0.163 \\ \hline
CrS2 & conduction & 3.3.497 & 0.9 & 3.025 & 0.920 & 0.50000 & 2.76148 & 86.682 & 3.511 & 2.658 & 0.527 & -2.052 \\ \hline
WOS & conduction & 3.3.289 & 1.51 & 2.987 & 0.567 & 8.26667 & 4.41192 & 89.937 & 5.020 & 4.496 & 0.828 & 2.176 \\ \hline
MoSTe & conduction & 3.3.286 & 1.03 & 3.340 & 0.780 & 0.11667 & 6.57796 & 89.010 & 2.028 & 2.546 & 0.757 & -4.813 \\ \hline
F2Ge & valence & 6.3.2420 & 2.64 & 3.563 & -1.683 & 1.25000 & 2.01197 & 69.229 & nan & nan & 0.174 & 1.095 \\ \hline
GaP & valence & 6.3.2065 & 1.55 & 3.881 & -0.658 & 95.66667 & 36.54672 & 64.063 & 3.046 & 1.221 & 0.846 & -13.376 \\ \hline
MoSeTe & conduction & 3.3.287 & 1.16 & 3.403 & 0.700 & 0.53333 & -4.21600 & 85.042 & 1.835 & 2.308 & 0.798 & -3.162 \\ \hline
CrO2 & conduction & 3.3.496 & 0.42 & 2.615 & 0.865 & 0.03333 & 0.92340 & 34.606 & 9.521 & 4.420 & 0.530 & -0.388 \\ \hline
CdPSe3 & valence & 3.3.315 & 1.27 & 6.526 & -0.491 & 7.56667 & -0.02250 & 0.000 & 2.586 & 0.936 & 0.043 & -0.007 \\ \hline
ZrIN & conduction & 6.3.2502 & 0.36 & 3.730 & 0.616 & 9.26667 & -0.43521 & 10.188 & 7.312 & 3.142 & 0.390 & -0.008 \\ \hline
ZnPSe3 & valence & 3.3.362 & 1.3 & 6.273 & -0.386 & 6.55000 & -0.16779 & 0.000 & 2.833 & 1.212 & 0.030 & 0.043 \\ \hline
ZrFN & conduction & 6.3.2427 & 2.32 & 3.553 & 0.567 & 2.30000 & 45.57193 & 86.781 & -4.528 & 1.212 & 2.013 & 11.153 \\ \hline
ZrFN & valence & 6.3.2427 & 2.32 & 3.553 & -1.128 & 0.31667 & 44.19712 & 86.814 & -4.528 & 1.212 & 2.013 & 11.153 \\ \hline
InP & valence & 6.3.2127 & 1.07 & 4.212 & -0.798 & 111.36667 & 53.22473 & 69.768 & 4.710 & 0.843 & 1.170 & -21.947 \\ \hline
GaAs & valence & 6.3.1940 & 1.07 & 4.027 & -0.586 & 139.36667 & 27.36001 & 59.109 & 2.283 & 1.000 & 0.902 & -3.824 \\ \hline
HgPS3 & valence & 3.3.334 & 0.98 & 6.308 & -0.660 & 6.63333 & 0.35167 & 0.000 & nan & nan & 0.043 & 0.021 \\ \hline
CrSe2 & conduction & 3.3.498 & 0.7 & 3.168 & 1.054 & 0.85000 & 3.16782 & 86.238 & 3.633 & 2.104 & 0.598 & -2.313 \\ \hline
AlN & valence & 6.3.2602 & 2.88 & 3.128 & -1.502 & 45.48333 & 90.32344 & 60.652 & 30.845 & 2.396 & 2.496 & -50.511 \\ \hline
MgAsSe3 & conduction & 3.3.308 & 1.28 & 6.458 & 0.374 & 21.10000 & 0.63151 & 0.000 & 1.280 & 1.218 & 0.035 & 0.020 \\ \hline
CrSSe & conduction & 3.3.276 & 0.8 & 3.096 & 1.024 & 0.20000 & -3.36467 & 88.230 & 3.511 & 2.361 & 0.558 & -2.398 \\ \hline
CdPS3 & valence & 3.3.314 & 1.91 & 6.240 & -0.645 & 5.91667 & 0.21411 & 0.000 & 3.282 & 1.051 & 0.035 & -0.012 \\ \hline
ZnPS3 & valence & 3.3.358 & 2.1 & 5.971 & -0.497 & 5.10000 & 0.51155 & 0.000 & 3.479 & 1.349 & 0.037 & 0.018 \\ \hline
MgPSe3 & valence & 3.3.343 & 2.01 & 6.357 & -0.521 & nan & nan & nan & 1.390 & 1.279 & 0.024 & -0.000 \\ \hline
HfFN & conduction & 6.3.2421 & 2.7 & 3.491 & 0.595 & 2.33333 & 35.00389 & 83.184 & -1.352 & 0.562 & 1.778 & 8.440 \\ \hline
HfFN & valence & 6.3.2421 & 2.7 & 3.491 & -1.159 & 0.55000 & 33.88649 & 83.436 & -1.352 & 0.562 & 1.778 & 8.440 \\ \hline
CrTe2 & conduction & 3.3.499 & 0.47 & 3.489 & 0.962 & 0.46667 & -3.60512 & 80.484 & 4.311 & 1.472 & 0.806 & -2.625 \\ \hline
AlGeS3 & conduction & 3.3.298 & 2.02 & 6.065 & 0.382 & 27.33333 & -0.18658 & 0.000 & 2.076 & 2.186 & 0.187 & -0.038 \\ \hline
InPt2Te3 & conduction & 6.3.2526 & 0.66 & 7.865 & 0.261 & 74.13333 & 12.08778 & 71.459 & 1.641 & 1.025 & 1.713 & -3.140 \\ \hline
MgAsS3 & conduction & 3.3.307 & 2.01 & 6.226 & 0.436 & 12.23333 & 0.24404 & 0.000 & 1.406 & 1.417 & 0.076 & -0.006 \\ \hline
HgPSe3 & valence & 3.3.335 & 0.56 & 6.593 & -0.489 & 8.38333 & -0.36370 & 0.000 & nan & nan & 0.061 & 0.102 \\ \hline
CrSeTe & conduction & 3.3.278 & 0.59 & 3.316 & 1.123 & 0.13333 & 4.54769 & 87.090 & 3.681 & 1.767 & 0.709 & -3.307 \\ \hline
YIS & conduction & 6.3.2506 & 2.56 & 4.455 & 1.750 & 7.50000 & 0.95465 & 83.138 & 12.188 & 0.799 & 0.367 & -0.201 \\ \hline
InPd2Te3 & conduction & 6.3.2522 & 0.37 & 7.855 & 0.276 & nan & nan & nan & 1.658 & 0.733 & 1.997 & -2.796 \\ \hline
CrSTe & conduction & 3.3.277 & 0.29 & 3.251 & 1.224 & 0.06667 & -3.90270 & 79.665 & 3.097 & 1.983 & 0.565 & -2.778 \\ \hline
MgPS3 & conduction & 3.3.342 & 2.81 & 6.089 & 0.429 & 8.63333 & -0.43499 & 0.000 & 1.502 & 1.519 & 0.033 & 0.017 \\ \hline
MgPS3 & valence & 3.3.342 & 2.81 & 6.089 & -0.674 & 4.35000 & 0.00946 & 0.000 & 1.502 & 1.519 & 0.033 & 0.017 \\ \hline
Pd2TlTe3 & conduction & 6.3.2568 & 0.37 & 7.857 & 0.331 & 74.60000 & 11.44748 & 75.545 & 1.241 & 0.745 & 1.638 & -1.951 \\ \hline
BN & conduction & 6.1.76 & 6.1 \cite{elias2019direct} & 2.514 & 0.937 & 106.81667 & 17.33421 & 81.688 & 3.500 & 7.800 & 0.270 & -7.248 \\ \hline
BN & valence & 6.1.76 & 6.1 \cite{elias2019direct} & 2.514 & -0.649 & 42.96667 & 13.21123 & 71.500 & 3.500 & 7.800 & 0.270 & -7.248 \\ \hline\hline
\end{longtable}

\begin{longtable}{c|c|c|c||c|c|c|c|c|c|c|c|c}
\caption{\textit{
Parameters for materials with spin-orbit coupling: $E/M_z$.
}}\label{tab_listparms_soc}\\
\toprule
Name & band & ID & $E_g$ [eV]& $a_0$ [$\AA$] & $\frac{m^*}{m_e}$ & $w$ [meV] & $v$ [meV] & $\psi$ [$^\circ$] & $\lambda$ $\left[\frac{\rm eV}{\AA^2}\right]$ & $\mu$ $\left[\frac{\rm eV}{\AA^2}\right]$ & $U_{\rm vdw}$ $\left[\frac{\rm meV}{\AA^2}\right]$ & $\delta_{\rm AB}$ $\left[\frac{\rm meV}{\AA}\right]$ \\[6pt]\hline
\midrule
\endfirsthead
\midrule
\endfoot
\bottomrule
\endlastfoot
WSe2 & conduction & 3.1.46 & 1.26 & 3.281 & 0.424 & 2.43333 & 3.43640 & 85.458 & 1.868 & 3.035 & 0.800 & -2.424 \\ \hline
WSe2 & valence & 3.1.46 & 1.26 & 3.281 & -0.386 & 9.23333 & -4.14754 & 89.385 & 1.868 & 3.035 & 0.800 & -2.424 \\ \hline
WS2 & conduction & 3.1.45 & 1.55 & 3.171 & 0.368 & 1.53333 & -2.98875 & 81.269 & 2.802 & 3.547 & 0.636 & -2.006 \\ \hline
WS2 & valence & 3.1.45 & 1.55 & 3.171 & -0.362 & 7.95000 & -3.36168 & 82.052 & 2.802 & 3.547 & 0.636 & -2.006 \\ \hline
MoS2 & valence & 3.1.39 & 1.6 & 3.164 & -0.558 & 4.83333 & -3.10387 & 88.860 & 3.073 & 3.072 & 0.619 & -1.961 \\ \hline
MoSe2 & valence & 3.1.41 & 1.34 & 3.282 & -0.632 & 6.35000 & -3.95509 & 89.370 & 2.197 & 2.623 & 0.751 & -2.574 \\ \hline
WTe2 & conduction & 3.1.47 & 0.75 & 3.524 & 0.387 & 2.90000 & 4.60916 & 83.586 & 1.188 & 2.315 & 0.864 & -3.468 \\ \hline
WTe2 & valence & 3.1.47 & 0.75 & 3.524 & -0.334 & 10.76667 & -6.00054 & 89.002 & 1.188 & 2.315 & 0.864 & -3.468 \\ \hline
MoTe2 & valence & 3.1.43 & 0.96 & 3.523 & -0.653 & 7.46667 & 5.15149 & 87.904 & 1.680 & 1.992 & 0.824 & -3.365 \\ \hline
MoSSe & valence & 3.1.10 & 1.48 & 3.223 & -0.639 & 3.30000 & -5.61730 & 88.268 & 2.548 & 2.850 & 0.711 & -3.684 \\ \hline
GaSe & valence & 6.2.1106 & 0.88 & 3.852 & -0.812 & nan & nan & nan & nan & nan & 3.152 & 3.044 \\ \hline
ZrBrN & conduction & 6.2.1099 & 2.59 & 3.476 & 0.533 & nan & nan & nan & nan & nan & 3.291 & -41.758 \\ \hline
HfCl2 & valence & 3.3.493 & 0.89 & 3.314 & -0.382 & 4.46667 & 1.39482 & 83.422 & 1.281 & 2.452 & 0.324 & 0.673 \\ \hline
CrS2 & valence & 3.3.497 & 0.9 & 3.025 & -0.892 & 2.86667 & -3.09417 & 86.633 & 3.511 & 2.658 & 0.530 & -2.006 \\ \hline
ZrBr2 & valence & 3.3.483 & 0.83 & 3.525 & -0.422 & 4.58333 & 2.24621 & 80.162 & 1.008 & 1.883 & 0.428 & -1.364 \\ \hline
WSSe & conduction & 3.3.292 & 1.42 & 3.226 & 0.435 & 1.55000 & 3.37022 & 84.281 & 2.213 & 3.298 & 0.618 & -2.391 \\ \hline
WSSe & valence & 3.3.292 & 1.42 & 3.226 & -0.403 & 4.45000 & 3.55570 & 89.422 & 2.213 & 3.298 & 0.618 & -2.391 \\ \hline
TiBr2 & valence & 3.3.482 & 0.76 & 3.434 & -0.599 & 3.56667 & 2.40171 & 86.051 & 1.281 & 1.811 & 0.400 & -1.514 \\ \hline
WSeTe & conduction & 3.3.294 & 1.06 & 3.400 & 0.569 & 3.38333 & -4.54664 & 89.912 & 1.385 & 2.697 & 0.856 & -3.279 \\ \hline
WSeTe & valence & 3.3.294 & 1.06 & 3.400 & -0.467 & 4.75000 & -4.72264 & 88.028 & 1.385 & 2.697 & 0.856 & -3.279 \\ \hline
HfBr2 & valence & 3.3.481 & 0.72 & 3.452 & -0.361 & 7.48333 & 2.74626 & 85.450 & 1.174 & 2.140 & 0.431 & -1.587 \\ \hline
ZrBrCl & valence & 3.3.267 & 0.91 & 3.460 & -0.443 & 5.48333 & 2.28056 & 89.731 & 1.100 & 2.007 & 0.418 & -1.610 \\ \hline
ZrI2 & valence & 3.3.507 & 0.7 & 3.809 & -0.375 & 7.03333 & 4.00815 & 87.085 & 0.701 & 1.580 & 0.509 & -2.496 \\ \hline
ZrClI & valence & 3.3.275 & 0.88 & 3.613 & -0.497 & 1.05000 & 0.62658 & 82.112 & 0.975 & 1.862 & 0.334 & -0.451 \\ \hline
WSTe & conduction & 3.3.293 & 1.17 & 3.342 & 0.907 & 0.61667 & 3.68541 & 65.676 & 1.603 & 2.987 & 0.726 & -4.342 \\ \hline
HfBrCl & valence & 3.3.265 & 0.82 & 3.388 & -0.384 & 9.01667 & 2.94259 & 87.451 & 1.194 & 2.279 & 0.420 & -1.880 \\ \hline
GeI2 & conduction & 6.3.2642 & 1.84 & 4.134 & 0.673 & 18.70000 & -2.87669 & 86.841 & nan & nan & 0.359 & -0.400 \\ \hline
HfI2 & valence & 3.3.505 & 0.62 & 3.738 & -0.293 & 11.61667 & 5.20685 & 88.836 & 0.857 & 1.728 & 0.514 & -3.043 \\ \hline
GaSe & conduction & 3.1.33 & 0.68 & 3.882 & 0.416 & nan & nan & nan & 0.903 & 0.854 & -6.857 & -9.764 \\ \hline
HfClI & valence & 3.3.273 & 0.81 & 3.538 & -0.431 & 1.75000 & 0.83279 & 88.138 & 1.033 & 2.091 & 0.340 & -1.009 \\ \hline
SnBrHO & valence & 6.3.2076 & 1.74 & 4.000 & -1.356 & 0.26667 & -0.47344 & 1.747 & nan & nan & 0.024 & 0.031 \\ \hline
TiI2 & valence & 3.3.506 & 0.6 & 3.753 & -0.511 & 4.43333 & 3.51899 & 88.088 & 1.061 & 1.508 & 0.495 & -2.255 \\ \hline
ZrBrI & valence & 3.3.270 & 0.78 & 3.671 & -0.430 & 2.75000 & -1.92761 & 86.744 & 0.838 & 1.737 & 0.444 & -1.364 \\ \hline
AsBrTe & conduction & 6.3.1931 & 1.1 & 3.841 & 0.553 & nan & nan & nan & 6.791 & 0.623 & 0.627 & -3.471 \\ \hline
TiClI & valence & 3.3.274 & 0.75 & 3.529 & -0.737 & 0.70000 & -0.97300 & 87.238 & 1.115 & 1.786 & 0.327 & -0.682 \\ \hline
AsClTe & conduction & 6.3.1936 & 1.32 & 3.810 & 0.636 & nan & nan & nan & -1.075 & 0.674 & 0.657 & -2.943 \\ \hline
MoSeTe & valence & 3.3.287 & 1.16 & 3.403 & -0.807 & 3.08333 & -4.63846 & 84.724 & 1.835 & 2.308 & 0.800 & -3.335 \\ \hline
HfBrI & valence & 3.3.268 & 0.7 & 3.600 & -0.361 & 4.41667 & 2.41585 & 88.725 & 0.938 & 1.935 & 0.452 & -1.661 \\ \hline
TiBrI & valence & 3.3.269 & 0.68 & 3.601 & -0.606 & 1.86667 & 2.26750 & 87.986 & 1.101 & 1.651 & 0.427 & -1.481 \\ \hline
HgF2 & valence & 6.3.2639 & 1.26 & 3.427 & -6.563 & 2.73333 & 1.63836 & 20.275 & nan & nan & 0.073 & 0.419 \\ \hline
CrSe2 & valence & 3.3.498 & 0.7 & 3.168 & -0.989 & 3.78333 & 3.55784 & 83.018 & 3.633 & 2.104 & 0.601 & -2.250 \\ \hline
BiO & conduction & 3.3.479 & 0.45 & 3.974 & 0.465 & 3.28333 & -12.31280 & 85.421 & -0.298 & 0.142 & 0.451 & 12.160 \\ \hline
CrSSe & valence & 3.3.276 & 0.8 & 3.096 & -1.040 & 1.73333 & -3.65244 & 88.879 & 3.511 & 2.361 & 0.560 & -2.441 \\ \hline
Sn2BrIS & valence & 6.3.2001 & 0.93 & 4.160 & -0.706 & 7.86667 & 0.72772 & 73.466 & 2.532 & 0.738 & 0.338 & -0.086 \\ \hline
PbTe & valence & 6.3.2560 & 0.73 & 5.503 & -2.801 & nan & nan & nan & nan & nan & 1.124 & -25.872 \\ \hline
IrBrSe & conduction & 6.3.2006 & 1.24 & 3.638 & 2.337 & nan & nan & nan & 1.175 & 1.506 & 1.077 & -3.289 \\ \hline
Hf3N2O2 & valence & 3.3.503 & 0.32 & 3.169 & -0.352 & nan & nan & nan & 9.488 & 9.736 & 1.278 & -11.275 \\ \hline
CrTe2 & valence & 3.3.499 & 0.47 & 3.489 & -0.923 & 4.06667 & -5.33941 & 87.858 & 4.311 & 1.472 & 0.816 & -3.353 \\ \hline
InSbSeTe & valence & 6.3.2128 & 0.34 & 4.686 & -0.081 & 2.25000 & -3.52371 & 86.471 & nan & nan & 0.804 & 0.591 \\ \hline
InPbBrSe & conduction & 6.3.2005 & 0.34 & 4.044 & 0.637 & nan & nan & nan & nan & nan & 0.590 & 0.233 \\ \hline
AsITe & conduction & 6.3.1945 & 0.42 & 3.958 & 0.514 & nan & nan & nan & 6.008 & 0.629 & 0.699 & -3.531 \\ \hline
AgAlP2Te6 & conduction & 3.3.215 & 0.52 & 6.870 & 0.727 & 16.81667 & 1.82601 & 76.472 & 6.528 & 0.750 & 0.074 & -0.508 \\ \hline
BiCuP2S6 & conduction & 6.3.1835 & 1.6 & 6.286 & 1.046 & 22.73333 & 2.81421 & 73.214 & 3.115 & 0.749 & 0.068 & -0.654 \\ \hline
AgP2S6Sb & conduction & 6.3.1824 & 1.69 & 6.357 & 0.838 & nan & nan & nan & -1.123 & 0.322 & 0.067 & -0.421 \\ \hline
CuP2S6Sb & conduction & 6.3.1839 & 1.69 & 6.245 & 0.667 & 25.00000 & 2.88748 & 71.459 & 5.767 & 0.627 & 0.068 & -0.542 \\ \hline
AuP2S6Sb & conduction & 6.3.1832 & 1.71 & 6.345 & 0.791 & nan & nan & nan & -2.887 & 0.391 & 0.036 & -0.247 \\ \hline
BiCl & valence & 1.3.200 & 0.91 & 5.351 & -0.272 & nan & nan & nan & -0.193 & 0.076 & 0.096 & -0.911 \\ \hline
BiBr & valence & 1.3.199 & 0.88 & 5.310 & -0.268 & nan & nan & nan & -0.140 & 0.059 & 0.373 & -1.802 \\ \hline
BiI & valence & 1.3.202 & 0.86 & 5.385 & -0.273 & nan & nan & nan & -0.100 & 0.045 & 0.408 & -0.248 \\ \hline
BrSb & conduction & 1.3.212 & 0.39 & 5.161 & 0.087 & nan & nan & nan & nan & nan & 0.188 & 0.031 \\ \hline
BrSb & valence & 1.3.212 & 0.39 & 5.161 & -0.074 & nan & nan & nan & nan & nan & 0.188 & 0.031 \\ \hline
ClSb & conduction & 1.3.254 & 0.39 & 5.180 & 0.088 & nan & nan & nan & nan & nan & 0.124 & -0.358 \\ \hline
ClSb & valence & 1.3.254 & 0.39 & 5.180 & -0.077 & nan & nan & nan & nan & nan & 0.124 & -0.358 \\ \hline
ReS & valence & 1.3.310 & 0.24 & 3.862 & -0.141 & nan & nan & nan & -30.211 & 1.904 & 0.559 & 2.432 \\ \hline
\end{longtable}

\begin{longtable}{c|c|c|c||c|c|c|c|c|c|c|c|c}
\caption{\textit{
Parameters for materials without spin-orbit coupling: $M_z/E$.
}}\label{tab_listparms}\\
\toprule
Name & band & ID & $E_g$ [eV]& $a_0$ [$\AA$] & $\frac{m^*}{m_e}$ & $w$ [meV] & $v$ [meV] & $\psi$ [$^\circ$] & $\lambda$ $\left[\frac{\rm eV}{\AA^2}\right]$ & $\mu$ $\left[\frac{\rm eV}{\AA^2}\right]$ & $U_{\rm vdw}$ $\left[\frac{\rm meV}{\AA^2}\right]$ & $\delta_{\rm AB}$ $\left[\frac{\rm meV}{\AA}\right]$ \\[6pt]\hline
\midrule
\endfirsthead
\midrule
\endfoot
\bottomrule
\endlastfoot
MoSSe & conduction & 3.1.10 & 1.48 & 3.223 & 0.560 & 2.41667 & 2.50006 & 78.392 & 2.548 & 2.850 & 0.705 & -2.010 \\ \hline
SnO & valence & 6.3.2139 & 1.68 & 3.338 & -1.043 & 8.58333 & 19.06411 & 88.477 & 1.698 & 0.804 & 0.934 & 11.100 \\ \hline
TiClF & valence & 3.3.271 & 1.13 & 3.075 & -0.985 & 0.23333 & 3.62110 & 81.315 & nan & nan & 0.670 & 2.414 \\ \hline
ZrClF & valence & 3.3.272 & 1.25 & 3.243 & -0.606 & 2.55000 & 2.42403 & 38.990 & nan & nan & 0.311 & -1.099 \\ \hline
ZrBr2O & conduction & 6.3.1978 & 1.07 & 3.662 & 0.618 & 10.03333 & 0.70998 & 44.810 & -13.611 & 1.397 & 0.328 & 0.159 \\ \hline
MoOS & conduction & 3.3.283 & 1.09 & 2.985 & 0.689 & 4.83333 & 2.03014 & 69.799 & 5.459 & 3.823 & 0.559 & -1.507 \\ \hline
WOSe & conduction & 3.3.290 & 1.26 & 3.043 & 0.903 & 10.36667 & 1.03257 & 26.870 & 3.933 & 4.242 & 0.671 & -1.115 \\ \hline
TiBrCl & valence & 3.3.266 & 0.83 & 3.353 & -0.654 & 1.53333 & 1.25765 & 82.234 & 1.343 & 1.953 & 0.316 & -0.811 \\ \hline
GeO & valence & 6.3.2066 & 2.09 & 2.992 & -1.226 & 8.05000 & 4.31405 & 87.805 & 1.926 & 1.104 & 0.425 & 2.350 \\ \hline
ISbTe & conduction & 6.3.2116 & 1.03 & 4.175 & 0.541 & 15.06667 & -2.67533 & 77.559 & -8.540 & 0.394 & 0.387 & -0.347 \\ \hline
MoOSe & conduction & 3.3.284 & 0.78 & 3.046 & 0.981 & 9.05000 & 0.97175 & 29.677 & 4.469 & 3.557 & 0.591 & -0.784 \\ \hline
Nb3F7S & conduction & 6.3.2051 & 0.96 & 6.155 & 3.791 & 13.68333 & 2.95359 & 45.652 & 2.540 & 1.455 & 0.223 & -0.516 \\ \hline
WOS & conduction & 3.3.289 & 1.51 & 2.987 & 0.567 & 3.76667 & 2.16363 & 72.547 & 5.020 & 4.496 & 0.569 & -1.525 \\ \hline
MoSTe & conduction & 3.3.286 & 1.03 & 3.340 & 0.780 & 5.88333 & 0.86318 & 24.711 & 2.028 & 2.546 & 0.701 & -1.002 \\ \hline
GaP & valence & 6.3.2065 & 1.55 & 3.881 & -0.658 & 63.56667 & 126.49828 & 63.467 & 3.046 & 1.221 & 1.881 & -54.152 \\ \hline
MoSeTe & conduction & 3.3.287 & 1.16 & 3.403 & 0.700 & 4.15000 & 1.46900 & 48.309 & 1.835 & 2.308 & 0.739 & -1.479 \\ \hline
InP & valence & 6.3.2127 & 1.07 & 4.212 & -0.798 & 64.20000 & 116.57425 & 65.020 & 4.710 & 0.843 & 2.213 & -51.244 \\ \hline
GaAs & valence & 6.3.1940 & 1.07 & 4.027 & -0.586 & 85.36667 & 134.95522 & 63.087 & 2.283 & 1.000 & 2.826 & -45.855 \\ \hline
CrSSe & conduction & 3.3.276 & 0.8 & 3.096 & 1.024 & 1.41667 & 1.41077 & 74.827 & 3.511 & 2.361 & 0.564 & -1.222 \\ \hline
CrSeTe & conduction & 3.3.278 & 0.59 & 3.316 & 1.123 & 1.85000 & 1.02162 & 68.088 & 3.681 & 1.767 & 0.628 & -0.943 \\ \hline
CrSTe & conduction & 3.3.277 & 0.29 & 3.251 & 1.224 & 2.86667 & 0.28650 & 67.464 & 3.097 & 1.983 & 0.592 & -0.514 \\ \hline\hline
\end{longtable}

\begin{longtable}{c|c|c|c||c|c|c|c|c|c|c|c|c}
\caption{\textit{
Parameters for materials with spin-orbit coupling: $M_z/E$.
}}\label{tab_listparms}\\
\toprule
Name & band & ID & $E_g$ [eV]& $a_0$ [$\AA$] & $\frac{m^*}{m_e}$ & $w$ [meV] & $v$ [meV] & $\psi$ [$^\circ$] & $\lambda$ $\left[\frac{\rm eV}{\AA^2}\right]$ & $\mu$ $\left[\frac{\rm eV}{\AA^2}\right]$ & $U_{\rm vdw}$ $\left[\frac{\rm meV}{\AA^2}\right]$ & $\delta_{\rm AB}$ $\left[\frac{\rm meV}{\AA}\right]$ \\[6pt]\hline
\midrule
\endfirsthead
\midrule
\endfoot
\bottomrule
\endlastfoot
MoSSe & valence & 3.1.10 & 1.48 & 3.223 & -0.639 & 8.06667 & 3.27428 & 85.043 & 2.548 & 2.850 & 0.708 & -1.815 \\ \hline
GaSe & valence & 6.2.1106 & 0.88 & 3.852 & -0.812 & 12.66667 & 1.64486 & 4.362 & nan & nan & 0.540 & -0.062 \\ \hline
ZrBrN & conduction & 6.2.1099 & 2.59 & 3.476 & 0.533 & 14.90000 & -0.80098 & 6.900 & nan & nan & 0.301 & 0.052 \\ \hline
WSSe & conduction & 3.3.292 & 1.42 & 3.226 & 0.435 & 1.05000 & 2.78808 & 85.889 & 2.213 & 3.298 & 0.773 & -1.913 \\ \hline
WSSe & valence & 3.3.292 & 1.42 & 3.226 & -0.403 & 13.35000 & -3.85402 & 85.907 & 2.213 & 3.298 & 0.773 & -1.913 \\ \hline
WSeTe & conduction & 3.3.294 & 1.06 & 3.400 & 0.569 & 0.68333 & 2.94180 & 75.892 & 1.385 & 2.697 & 0.767 & -2.246 \\ \hline
WSeTe & valence & 3.3.294 & 1.06 & 3.400 & -0.467 & 20.36667 & -5.87390 & 86.072 & 1.385 & 2.697 & 0.767 & -2.246 \\ \hline
ZrBrCl & valence & 3.3.267 & 0.91 & 3.460 & -0.443 & 2.00000 & 1.02998 & 77.497 & 1.100 & 2.007 & 0.325 & -0.606 \\ \hline
ZrClI & valence & 3.3.275 & 0.88 & 3.613 & -0.497 & 11.20000 & 3.94119 & 88.862 & 0.975 & 1.862 & 0.504 & -2.288 \\ \hline
WSTe & conduction & 3.3.293 & 1.17 & 3.342 & 0.907 & 1.98333 & 2.13545 & 61.756 & 1.603 & 2.987 & 0.758 & -1.195 \\ \hline
HfBrCl & valence & 3.3.265 & 0.82 & 3.388 & -0.384 & 3.46667 & 1.20943 & 75.169 & 1.194 & 2.279 & 0.332 & -0.924 \\ \hline
HfClI & valence & 3.3.273 & 0.81 & 3.538 & -0.431 & 18.26667 & 5.51265 & 86.006 & 1.033 & 2.091 & 0.511 & -3.007 \\ \hline
SnBrHO & valence & 6.3.2076 & 1.74 & 4.000 & -1.356 & 14.75000 & -2.61215 & 37.698 & nan & nan & 0.306 & 0.047 \\ \hline
ZrBrI & valence & 3.3.270 & 0.78 & 3.671 & -0.430 & 10.06667 & 3.74995 & 86.550 & 0.838 & 1.737 & 0.507 & -2.287 \\ \hline
AsBrTe & conduction & 6.3.1931 & 1.1 & 3.841 & 0.553 & nan & nan & nan & 6.791 & 0.623 & 0.306 & -0.366 \\ \hline
TiClI & valence & 3.3.274 & 0.75 & 3.529 & -0.737 & 8.91667 & 3.36966 & 86.415 & 1.115 & 1.786 & 0.486 & -1.819 \\ \hline
AsClTe & conduction & 6.3.1936 & 1.32 & 3.810 & 0.636 & nan & nan & nan & -1.075 & 0.674 & 0.225 & 1.117 \\ \hline
MoSeTe & valence & 3.3.287 & 1.16 & 3.403 & -0.807 & 11.76667 & -3.31038 & 89.307 & 1.835 & 2.308 & 0.747 & -1.618 \\ \hline
HfBrI & valence & 3.3.268 & 0.7 & 3.600 & -0.361 & 16.53333 & 5.34105 & 89.174 & 0.938 & 1.935 & 0.515 & -2.973 \\ \hline
TiBrI & valence & 3.3.269 & 0.68 & 3.601 & -0.606 & 7.35000 & 3.34909 & 89.047 & 1.101 & 1.651 & 0.489 & -1.954 \\ \hline
CrSSe & valence & 3.3.276 & 0.8 & 3.096 & -1.040 & 5.46667 & 1.89848 & 82.265 & 3.511 & 2.361 & 0.566 & -1.164 \\ \hline
Sn2BrIS & valence & 6.3.2001 & 0.93 & 4.160 & -0.706 & 12.76667 & -4.49953 & 8.238 & 2.532 & 0.738 & 0.353 & 0.336 \\ \hline
IrBrSe & conduction & 6.3.2006 & 1.24 & 3.638 & 2.337 & nan & nan & nan & 1.175 & 1.506 & 0.397 & 0.155 \\ \hline
InSbSeTe & valence & 6.3.2128 & 0.34 & 4.686 & -0.081 & nan & nan & nan & nan & nan & 0.717 & -3.470 \\ \hline
InPbBrSe & conduction & 6.3.2005 & 0.34 & 4.044 & 0.637 & 47.00000 & 1.38099 & 7.205 & nan & nan & 0.355 & 0.034 \\ \hline
AsITe & conduction & 6.3.1945 & 0.42 & 3.958 & 0.514 & 10.61667 & 4.12550 & 88.105 & 6.008 & 0.629 & 0.386 & 0.526 \\ \hline
BiCuP2S6 & conduction & 6.3.1835 & 1.6 & 6.286 & 1.046 & 23.23333 & 2.08464 & 77.492 & 3.115 & 0.749 & 0.053 & -0.509 \\ \hline
AgP2S6Sb & conduction & 6.3.1824 & 1.69 & 6.357 & 0.838 & nan & nan & nan & -1.123 & 0.322 & 0.050 & -0.302 \\ \hline
CuP2S6Sb & conduction & 6.3.1839 & 1.69 & 6.245 & 0.667 & nan & nan & nan & 5.767 & 0.627 & 0.048 & -0.476 \\ \hline
AuP2S6Sb & conduction & 6.3.1832 & 1.71 & 6.345 & 0.791 & nan & nan & nan & -2.887 & 0.391 & 0.050 & -0.440 \\ \hline\hline
\end{longtable}

\begin{longtable}{c|c|c|c}
\caption{\textit{
Atomic configuration of $E/M_z$ and $M_z/E$ without spin-orbit coupling.
}}\label{tab_liststack}\\
\toprule
Name & ID & Arrangement & Atomic order along Z-axis \\[6pt]\hline
\midrule
\endfirsthead
\midrule
\endfoot
\bottomrule
\endlastfoot

MoSSe & 3.1.10 & $E/M_z$ & Se-Mo-S   S-Mo-Se \\ \hline
  &  & $M_z/E$ & S-Mo-Se   Se-Mo-S \\ \hline
SnO & 6.3.2139 & $E/M_z$ & O-Sn   Sn-O \\ \hline
  &  & $M_z/E$ & Sn-O   O-Sn \\ \hline
TiClF & 3.3.271 & $E/M_z$ & F-Ti-Cl   Cl-Ti-F \\ \hline
  &  & $M_z/E$ & Cl-Ti-F   F-Ti-Cl \\ \hline
ZrClF & 3.3.272 & $E/M_z$ & Cl-Zr-F   F-Zr-Cl \\ \hline
  &  & $M_z/E$ & F-Zr-Cl   Cl-Zr-F \\ \hline
ZrBr2O & 6.3.1978 & $E/M_z$ & Br-Zr-O-Br   Br-O-Zr-Br \\ \hline
  &  & $M_z/E$ & Br-O-Zr-Br   Br-Zr-O-Br \\ \hline
MoOS & 3.3.283 & $E/M_z$ & S-Mo-O   O-Mo-S \\ \hline
  &  & $M_z/E$ & O-Mo-S   S-Mo-O \\ \hline
WOSe & 3.3.290 & $E/M_z$ & Se-W-O   O-W-Se \\ \hline
  &  & $M_z/E$ & O-W-Se   Se-W-O \\ \hline
TiBrCl & 3.3.266 & $E/M_z$ & Cl-Ti-Br   Br-Ti-Cl \\ \hline
  &  & $M_z/E$ & Br-Ti-Cl   Cl-Ti-Br \\ \hline
GeO & 6.3.2066 & $E/M_z$ & O-Ge   Ge-O \\ \hline
  &  & $M_z/E$ & Ge-O   O-Ge \\ \hline
ISbTe & 6.3.2116 & $E/M_z$ & I-Sb-Te   Te-Sb-I \\ \hline
  &  & $M_z/E$ & Te-Sb-I   I-Sb-Te \\ \hline
MoOSe & 3.3.284 & $E/M_z$ & Se-Mo-O   O-Mo-Se \\ \hline
  &  & $M_z/E$ & O-Mo-Se   Se-Mo-O \\ \hline
Nb3F7S & 6.3.2051 & $E/M_z$ & S-F3-Nb3-F4   F4-Nb3-F3-S \\ \hline
  &  & $M_z/E$ & F4-Nb3-F3-S   S-F3-Nb3-F4 \\ \hline
WOS & 3.3.289 & $E/M_z$ & S-W-O   O-W-S \\ \hline
  &  & $M_z/E$ & O-W-S   S-W-O \\ \hline
MoSTe & 3.3.286 & $E/M_z$ & Te-Mo-S   S-Mo-Te \\ \hline
  &  & $M_z/E$ & S-Mo-Te   Te-Mo-S \\ \hline
GaP & 6.3.2065 & $E/M_z$ & Ga-P   P-Ga \\ \hline
  &  & $M_z/E$ & P-Ga   Ga-P \\ \hline
MoSeTe & 3.3.287 & $E/M_z$ & Te-Mo-Se   Se-Mo-Te \\ \hline
  &  & $M_z/E$ & Se-Mo-Te   Te-Mo-Se \\ \hline
InP & 6.3.2127 & $E/M_z$ & In-P   P-In \\ \hline
  &  & $M_z/E$ & P-In   In-P \\ \hline
GaAs & 6.3.1940 & $E/M_z$ & Ga-As   As-Ga \\ \hline
  &  & $M_z/E$ & As-Ga   Ga-As \\ \hline
CrSSe & 3.3.276 & $E/M_z$ & Se-Cr-S   S-Cr-Se \\ \hline
  &  & $M_z/E$ & S-Cr-Se   Se-Cr-S \\ \hline
CrSeTe & 3.3.278 & $E/M_z$ & Te-Cr-Se   Se-Cr-Te \\ \hline
  &  & $M_z/E$ & Se-Cr-Te   Te-Cr-Se \\ \hline
CrSTe & 3.3.277 & $E/M_z$ & Te-Cr-S   S-Cr-Te \\ \hline
  &  & $M_z/E$ & S-Cr-Te   Te-Cr-S \\ \hline\hline
\end{longtable}

\begin{longtable}{c|c|c|c}
\caption{\textit{
Atomic configuration of $E/M_z$ and $M_z/E$ with spin-orbit coupling.
}}\label{tab_liststacksoc}\\
\toprule
Name & ID & Arrangement & Atomic order along Z-axis \\[6pt]\hline
\midrule
\endfirsthead
\midrule
\endfoot
\bottomrule
\endlastfoot

MoSSe & 3.1.10 & $E/M_z$ & Se-Mo-S   S-Mo-Se \\ \hline
  &  & $M_z/E$ & S-Mo-Se   Se-Mo-S \\ \hline
GaSe & 6.2.1106 & $E/M_z$ & Se-Ga-Se-Ga   Ga-Se-Ga-Se \\ \hline
  &  & $M_z/E$ & Ga-Se-Ga-Se   Se-Ga-Se-Ga \\ \hline
ZrBrN & 6.2.1099 & $E/M_z$ & Br-Zr-N   N-Zr-Br \\ \hline
  &  & $M_z/E$ & N-Zr-Br   Br-Zr-N \\ \hline
WSSe & 3.3.292 & $E/M_z$ & Se-W-S   S-W-Se \\ \hline
  &  & $M_z/E$ & S-W-Se   Se-W-S \\ \hline
WSeTe & 3.3.294 & $E/M_z$ & Te-W-Se   Se-W-Te \\ \hline
  &  & $M_z/E$ & Se-W-Te   Te-W-Se \\ \hline
ZrBrCl & 3.3.267 & $E/M_z$ & Cl-Zr-Br   Br-Zr-Cl \\ \hline
  &  & $M_z/E$ & Br-Zr-Cl   Cl-Zr-Br \\ \hline
ZrClI & 3.3.275 & $E/M_z$ & I-Zr-Cl   Cl-Zr-I \\ \hline
  &  & $M_z/E$ & Cl-Zr-I   I-Zr-Cl \\ \hline
WSTe & 3.3.293 & $E/M_z$ & Te-W-S   S-W-Te \\ \hline
  &  & $M_z/E$ & S-W-Te   Te-W-S \\ \hline
HfBrCl & 3.3.265 & $E/M_z$ & Cl-Hf-Br   Br-Hf-Cl \\ \hline
  &  & $M_z/E$ & Br-Hf-Cl   Cl-Hf-Br \\ \hline
HfClI & 3.3.273 & $E/M_z$ & I-Hf-Cl   Cl-Hf-I \\ \hline
  &  & $M_z/E$ & Cl-Hf-I   I-Hf-Cl \\ \hline
SnBrHO & 6.3.2076 & $E/M_z$ & Br-Sn-O-H   H-O-Sn-Br \\ \hline
  &  & $M_z/E$ & H-O-Sn-Br   Br-Sn-O-H \\ \hline
ZrBrI & 3.3.270 & $E/M_z$ & I-Zr-Br   Br-Zr-I \\ \hline
  &  & $M_z/E$ & Br-Zr-I   I-Zr-Br \\ \hline
AsBrTe & 6.3.1931 & $E/M_z$ & Br-As-Te   Te-As-Br \\ \hline
  &  & $M_z/E$ & Te-As-Br   Br-As-Te \\ \hline
TiClI & 3.3.274 & $E/M_z$ & I-Ti-Cl   Cl-Ti-I \\ \hline
  &  & $M_z/E$ & Cl-Ti-I   I-Ti-Cl \\ \hline
AsClTe & 6.3.1936 & $E/M_z$ & Cl-As-Te   Te-As-Cl \\ \hline
  &  & $M_z/E$ & Te-As-Cl   Cl-As-Te \\ \hline
MoSeTe & 3.3.287 & $E/M_z$ & Te-Mo-Se   Se-Mo-Te \\ \hline
  &  & $M_z/E$ & Se-Mo-Te   Te-Mo-Se \\ \hline
HfBrI & 3.3.268 & $E/M_z$ & I-Hf-Br   Br-Hf-I \\ \hline
  &  & $M_z/E$ & Br-Hf-I   I-Hf-Br \\ \hline
TiBrI & 3.3.269 & $E/M_z$ & I-Ti-Br   Br-Ti-I \\ \hline
  &  & $M_z/E$ & Br-Ti-I   I-Ti-Br \\ \hline
CrSSe & 3.3.276 & $E/M_z$ & Se-Cr-S   S-Cr-Se \\ \hline
  &  & $M_z/E$ & S-Cr-Se   Se-Cr-S \\ \hline
Sn2BrIS & 6.3.2001 & $E/M_z$ & I-Sn-S-Sn-Br   Br-Sn-S-Sn-I \\ \hline
  &  & $M_z/E$ & Br-Sn-S-Sn-I   I-Sn-S-Sn-Br \\ \hline
IrBrSe & 6.3.2006 & $E/M_z$ & Br-Ir-Se   Se-Ir-Br \\ \hline
  &  & $M_z/E$ & Se-Ir-Br   Br-Ir-Se \\ \hline
InSbSeTe & 6.3.2128 & $E/M_z$ & Se-Sb-In-Te   Te-In-Sb-Se \\ \hline
  &  & $M_z/E$ & Te-In-Sb-Se   Se-Sb-In-Te \\ \hline
InPbBrSe & 6.3.2005 & $E/M_z$ & Br-Pb-In-Se   Se-In-Pb-Br \\ \hline
  &  & $M_z/E$ & Se-In-Pb-Br   Br-Pb-In-Se \\ \hline
AsITe & 6.3.1945 & $E/M_z$ & I-As-Te   Te-As-I \\ \hline
  &  & $M_z/E$ & Te-As-I   I-As-Te \\ \hline
RhISe & 6.3.2109 & $E/M_z$ & I-Rh-Se   Se-Rh-I \\ \hline
  &  & $M_z/E$ & Se-Rh-I   I-Rh-Se \\ \hline
BiCuP2S6 & 6.3.1835 & $E/M_z$ & S3-P-Bi-P-Cu-S3   S3-Cu-P-Bi-P-S3 \\ \hline
  &  & $M_z/E$ & S3-Cu-P-Bi-P-S3   S3-P-Bi-P-Cu-S3 \\ \hline
AgP2S6Sb & 6.3.1824 & $E/M_z$ & S3-P-Sb-Ag-P-S3   S3-P-Ag-Sb-P-S3 \\ \hline
  &  & $M_z/E$ & S3-P-Ag-Sb-P-S3   S3-P-Sb-Ag-P-S3 \\ \hline
CuP2S6Sb & 6.3.1839 & $E/M_z$ & S3-P-Sb-P-Cu-S3   S3-Cu-P-Sb-P-S3 \\ \hline
  &  & $M_z/E$ & S3-Cu-P-Sb-P-S3   S3-P-Sb-P-Cu-S3 \\ \hline
AuP2S6Sb & 6.3.1832 & $E/M_z$ & S3-P-Sb-P-Au-S3   S3-Au-P-Sb-P-S3 \\ \hline
  &  & $M_z/E$ & S3-Au-P-Sb-P-S3   S3-P-Sb-P-Au-S3 \\ \hline\hline
\end{longtable}

\end{document}